%% file: main.tex
\documentclass[letterpaper,english,aps,prb,floatfix,twocolumn,amsfonts,superscriptaddress,longbibliography]{revtex4-1}
\input{Laras_tex_library}

\usepackage[T1]{fontenc}
\usepackage[applemac]{inputenc}
\usepackage[english]{babel}
\usepackage{graphics}
\usepackage{amssymb}
\usepackage{amsmath}
\usepackage{overpic}
\usepackage{comment}
\usepackage{bbold}
\usepackage{siunitx}
\usepackage[normalem]{ulem}
\usepackage{natbib}

\usepackage{xcolor}
\usepackage{soul}

\graphicspath{{./Images/}}

\def\mathO{{\mathcal O}}

\def\Tr{{\mathrm Tr}}
\newcommand{\clg}[1]{ {\color{green} #1}}

\begin{document}
	\title{Boltzmann electronic dc transport in multiorbital weakly-disordered crystals}
	\author{Marco Marciani}
	\affiliation{Department of Physics and ISC-CNR, ``Sapienza'' University of Rome, P.le A. Moro 5, 00185 Rome, Italy}
	\author{Lara Benfatto}
	\affiliation{Department of Physics and ISC-CNR, ``Sapienza'' University of Rome, P.le A. Moro 5, 00185 Rome, Italy}
	
	\begin{abstract}
		Motivated by the increasing number of systems featuring multiple bands at low energy, we address the Boltzmann approach to transport in a multiband weakly disordered noninteracting crystal subject to a small electric field. In general, the multiband structure leads to a considerable complication of the Boltzmann equation. Indeed, even in the presence of elastic impurity scattering, one needs to compute for each band and momentum the dressed velocities, which account for scattering events. Here we provide a semianalytical solution to the Boltzmann equation that reduces such a challenging numerical task to the much simpler numerical computation of a small tensor whose dimension is set by the number of bands at the Fermi level. This approach further allows us to discuss the interplay of symmetry and disorder for different impurity types, including those originating from random-matrix Wigner ensembles. As an example of application, we consider the 2D isotropic Rashba metal and we discuss, in a full analytical fashion, how different types of disorders may break the exactness of the relaxation-time approximation and induce transport anisotropy, and may allow one to identify the presence of spin-orbit coupling as deviations of the conductivity from the Drude behavior.
	\end{abstract}
	\date{\today}

	\maketitle
	
	\section{Introduction}

	In the last decades, due to enhanced capabilities in materials fabrication new classes of materials have been crafted whose electronic transport properties are challenging to model as several degrees of freedom and atomic orbitals take part in the low-energy physics\cite{Das2011,Gar16,Wang2017,Man17,Fer16,Gor2018, San2018, Gil21}. 
	The effect of the scattering in the transport can be usually computed semianalytically by means of either the Boltzmann approach\cite{Ziman60,Mah00} or some kind of diagrammatic Green's functions expansion (Matsubara\cite{Mah00}, Keldysh\cite{Kam2011}, supersymmetry or replica trick\cite{Efe97}) and new techniques have been invented recently relying on quantum master equations\cite{Cul2017,Sek2017,Cong19,Sted20,Cep21} or holographic dualities\cite{Fau10,Ada15,Luc15}. Within the first approach, the attention is often focused on the difference between quasiparticle and the transport lifetimes\cite{Mah00, Sch02}, with the idea that while evaluating the transport the scattering events which do not change the direction of the electronic velocity do not affect the current. Such a difference is easily seen by solving the problem for a single-band isotropic system, where the dc conductivity relating the dc electric field and the induced current $j_x= \sigma_{xx} E_x$ reads
	\be
	\lb{sigma_intro}
	\sigma_{xx}=2e^2\sum_\bk v_{\bk,x}^2 \tau^{tr}_\bk \left(-\pd f/\pd \e_\bk\right)
	\ee
	where $e$ is the electronic charge, $v_{\bk,x}$ is the band velocity at momentum $\bk$ and $f$ is the Fermi equilibrium distribution. The quantity $1/\tau^{tr}_\bk=\sum_{\bk'}Q_{\bk\bk'}(1-\cos\theta_{\bk-\bk'})$
	is the inverse transport scattering time which is to be compared to $1/\tau_\bk=\sum_{\bk'}Q_{\bk,\bk'}$, the inverse quasiparticle lifetime. Whenever the scattering kernel $Q_{\bk\bk'}$ is momentum independent, as it occurs for instance for scattering off impurities localized in the unit cell, $\tau^{tr}_{\bk}$ coincides with $\tau_{\bk}$ and the transport is trivial. 
	At a more general level, the difference between the two lifetimes should be rephrased as the difference between the {\it bare} electronic velocities $\bv_\bk\equiv \pd \e_\bk/\pd \bk$, which are simply determined by the band dispersion, and the {\it dressed} 
	ones ${\bf w}_\bk$, which enter the generalization of Eq.\ \pref{sigma_intro} to arbitrary band curvature:
	\be
	\sigma_{xx}= 2e^2\sum_\bk v_{\bk,x}  w_{\bk,x} \tau_\bk \left(-\frac{\pd f}{\pd \e_\bk}\right)
	\ee
	
	This language makes it easier to put in correspondence Boltzmann theory and diagrammatic approaches, where the difference between ${\bf w}$ and $\bv$ is determined by a summation of  ladder diagrams, involving the impurities propagators, at one of the two current vertices constituting the diagram for the conductivity, under Lorentzian approximation of the electronic Green's functions\cite{Mah00,Sch02,Bros16}.
	
	Stricly speaking, transport lifetimes may be defined only when all ${\bf w}_\bk$ and $\bv_\bk$ are parallel, which happens only for highly symmetric systems. However if this is not the case an approximation is often made whereby parallelism is enforced, known as relaxation time approximation (RTA)\cite{Mah00}. 
	
	In multiband systems things are worse. Isotropy of bands and interactions are not sufficient conditions for RTA neither to be exact nor even to be a good approximation, since also the multiorbital character of the eigenstates and, if included, of the disorder enter the game. Therefore, within the Boltzmann approach it is of interest to be able to compute properly ${\bf w}_\bk$, a task which requires to go beyond RTA and calls for a full numerical treatment, unless other approximations are done\cite{All78,Bre14}.
	
	In this paper we show that an analytical solution exists for multiorbital systems in the case of scattering induced by weak and localized disorder which possibly mixes the orbital degrees of freedoms. Importantly, it is possible to understand the different roles of the multiorbital nature of the Bloch eigenstates and of the impurity type in determining the current. Symmetries do play an important role in determining whether vertex corrections vanish in a system (i.e. $\bf w_\bk = v_\bk$) or, more generically, whether the RTA happens to be exact.
	
	 As an application we focus our attention a 2D Rashba electron gas\cite{Byc84}. Such model has been experimentally realized in a number of systems such as HgTe quantum wells\cite{Gui04}, layered bismuth tellurohalides\cite{Ere12,Bah12} and interfaces between complex oxides\cite{Oht04,Cav08}. In Ref. \onlinecite{Bros16} the model was solved assuming simple intraorbital disorder and it was discovered that the Rashba physics manifests itself as a departure of the conductivity formula from the Drude one\cite{Mah00}. Within our approach we compute fully analytically the transport quantities, discuss the exactness of the RTA and go beyond that work by showing how the different disorder types lead to qualitatively different results. In particular, with a certain magnetic disorder the Rasbha conductivity may not feature any departure from the Drude one.

	\section{The Boltzmann equation -- statement of the problem}
	
	We are interested in investigating metals or semiconductors where the electronic bands close to the Fermi surface display multiorbital character. Let the index $b$ label the electronic bands, then we define $\rho^b_{\bf k}$ 
	as the band-resolved electron occupancy at the momentum $\bf k$ in the 3D 
	Brillouin zone (BZ). If we aim at describing the linear static response of the material, we are interested in finding the non-equilibrium time-independent electronic configuration at small external electric field $\bf E$. Then the Boltzmann equation reads
	\begin{equation} \label{boltz_eq}
		e\, {\bf E}\cdot \nabla_{\bf k} \, \rho^b_{\bf r, \bf k}  =  \sum_{\bk',b'} Q_{\bk,b}^{\bk',b'}(\rho^b_{\bf r, \bf k} - \rho^{b'}_{\bf r, \bk'}),\quad 0\leq b \leq N_{b}.  
	\end{equation}
	The equation states that at equilibrium the push from the electric field on the momentum of an electron in a specific band is counteracted exactly by diffusion processes, mixing states of all bands and momenta. The matrix $Q_{\bf k,b}^{\bf k',b'}$ regulates such diffusion and is referred to as the collision-integral kernel. For what concerns us it will include only scattering from onsite disorder due to impurities and, as usually done, we write it according to the Fermi golden's rule, leading to a treatment of disorder equivalent to the Born approximation within the Green's function formalism\cite{Mah00}:
	\begin{equation} \lb{Qmulti1}
		Q^{bb'}_{\bf k k'} = \frac{2\pi }{N_L}\left\vert \langle {\bf k},b \vert \sum_{i}^{N_L} {\vec c}_i^{\;\dagger} \cdot V_i \cdot\,{\vec c}_{i}  \;\vert \bk',b' \rangle \right\vert^2 \delta(\xi_{{\bf k},b}-\xi_{{\bf k'},b'}).
	\end{equation}
	Here $\vert \bk, b\rangle $ denotes the Bloch state for the $b$ band, the 
	index $i$ labels lattice sites, $N_L$ is the number of sites, and $V_i$ the impurity matrix at site $i$ written in the original orbital basis, such that the $l$-th element of the ${\vec c}_i$ annihilates the $l$-th orbital state at site $i$. Hereafter we set $\hbar=1$. Notice that the kernel couples only states on the same energy shell, since in this model the impurities do not have internal degrees of freedom to absorb energy. In the thermodynamic limit, $N_L \rightarrow \infty$, the matrix $Q$ becomes an operator and we will refer to it with one or the other term interchangeably.
	The quasiparticle scattering rates $\G_{\bk,b}$ and lifetimes $\t_{\bk,b} $ are defined as
	\be
	\lb{gkb}
	\G_{\bk,b} = 1/\t_{\bk,b} = \sum_{\bk',b'} Q^{bb'}_{\bf k \bf k'}.
	\ee

	At small field $\bf E$ we can set $\rho_{\bk,b} = f_{\varepsilon^b_{\bf 
			k}} + \rho^{\bf E}_{\bk,b}$ where $f_{\varepsilon^b_{\bf k}}$ is the temperature-dependent equilibrium Fermi distribution while the second term is 
	further parametrized through a vector ${\bf w}_{\bk,b}$ as
	\be \lb{pop_corr}
	\rho^{\bf E}_{\bk,b} = e {\bf E} \cdot {\bf w}_{\bk,b} \,\t_{\bk,b} \,\partial_{\varepsilon^b_{\bf k}} f_{\varepsilon^b_{\bf k}}
	\ee
	with the notation $\partial_{\varepsilon^b_{\bf k}}:=\partial/\partial_{\varepsilon^b_{\bf k}}$. This term is linear in the field and is the population correction we ought to find. The vector $\bf w$ is proportional to the first-order Born-approximated current dressed with vertex corrections (with the additional Lorentzian approximation of the internal Green's functions) as computed within a fully-quantum diagrammatic approach\cite{Mah00,Sch02,Bros16}. 
	
	The Boltzmann equation at zeroth-order in $\bf E$ is trivially solved by $f_{\varepsilon^b_{\bf k}}$ as it nullifies the r.h.s of Eq. \ref{boltz_eq}. Instead the first-order equation leads to the sought equation for ${\bf w}_{\bk,b}$:
	\begin{eqnarray} \lb{w_solver}
		\sum_{\bf k',b'}\left(\delta_{\bf k\bf k'}\delta_{bb'} - Q^{bb'}_{\bf k\bf k'}\t^{b'}_{\bf k'}\right) {\bf w}^b_{\bf k'} =  {\bf v}^b_{\bf k}
	\end{eqnarray}
	where we define the band velocities as ${\bf v}^b_{\bf k}= \nabla_{\bf k}\varepsilon^b_{\bf k}$. 
	
	Computing the population corrections amounts to 
	inverting the matrix $1-Q\t$ in the grouped indexes $(\bk b)$ and $(\bk'b')$ in the l.h.s. of the equation \eqref{w_solver}. Unfortunately it is impossible to perform this task exactly in generic systems. Often the RTA\cite{Mah00} is invoked, which in essence turn the operator $Q$ into a vector. This approximation might even be exact in some highly symmetric systems\cite{Bros16}(see Sec. \ref{sec:Rashba}) but is unreliable in most cases. At best an expansion of $Q$ in terms of orthogonal polynomials may be performed and truncated to have a treatable finite-rank matrix\cite{All78}. In this manuscript we show that in the case of elastic scattering by impurities a finite-rank matrix can be spotted without making approximations and the problem can be enormously simplified. Indeed, the energy conservation in the collision-integral kernel makes 
	the matrix $1-Q\t$ block-diagonal when the $\bk$ vectors are ordered in groups belonging to the same energy shell. We will show that these blocks are finite-rank thus allowing to reduce the problem to the inversion of a "small" matrix whose size $N^2_b$ is set by number of orbitals.
	
	Before jumping on the solution of the multiband case, it is instructive to present the solution for the single-band problem.

	\section{Single-band solution}
	With a single band we can safely remove band labels in the previous expressions. The square modulus in Eq. \ref{Qmulti1} can be expressed as a sum 
	in space: $\left\vert \langle {\bf k}\vert \sum_{i}^{N_L} c_i^{\;\dagger} 
	\cdot V_i \cdot\,c_{i}  \;\vert \bk' \rangle \right\vert^2 = \frac{1}{N_L^2}\sum_{jl}^{N_L} V_j V_l e^{i(\bx_j-\bx_l)\cdot(\bk-\bk')}$, where we 
	used $c_j =\sum_\bk c_\bk e^{i\bx_j\cdot\bk}/\sqrt{N_L}$. 
	Let's assume now that the disorder is made of $N_I$ impurities acting only onsite with strength $v_I$ and whose locations are uniformly random in the sample and uncorrelated to each other i.e.
	\begin{equation} \label{Vi}
		V_i=v_I \sum_{c}^{N_I} \delta_{i j_c}
	\end{equation}
	with $j_c$ the random location of an impurity. Then, in the summation considered above, the product of the impurity potential at different sites self-averages\cite{Mah00,Koh57,Lut58} in the thermodynamic limit i.e. $V_j V_l \rightarrow \langle V_j V_l\rangle_V \simeq N_I v^2_{I} \delta_{jl}$, where $\langle\cdot\rangle_V =\frac{1}{N_L^{N_I}}\prod_{c}^{N_I}\sum_{j_c}^{N_L} (\cdot)$ denotes the average over spatial locations and the last equality holds in the limit $n_I = N_I/N_L \ll 1$.
	Within this model the collision integral is a function of the energies only:
	\begin{equation} 
		\lb{defkappa}
		Q_{\bf k k'} = \kappa\;\delta(\xi_{\bf k}-\xi_{\bf k'}), \quad \kappa = 2\pi n_I v^2_{I}/N_L
	\end{equation}
	It follows that the rates (and the lifetimes) are also energy functions: 
	\begin{equation} \label{G_single}
	    \G^{(single)}_{\bf k} =  \kappa\, \mathcal{N}(\varepsilon_{\bf k}),
	\end{equation}
	where $\mathcal{N}(\varepsilon)$ denotes the total density of states per unit energy. At this point it is straightforward to see that the solution of Eq. \pref{w_solver} is trivial: $\bf w_{\bf k} = \bf v_{\bf k}$ i.e. 
	no dressing of the current occurs in the single-band case. In principle we may add to the solution any function of the energy, but as will be argued in the next section consistency requirements imply that this function must be identically zero.
	
	We note  that this result is due to the specific isotropic disorder at hand. Had $\kappa$ not been a constant function of momenta then there would 
	have been some dressing. As we shall see in a moment, this is exactly what will happen in the multiband case. 
	
	\section{multiband solution}
	Multiband systems feature a difference between bare and dressed velocities even when the disorder potential is isotropic. Two new ingredients add to the equations as compared to single-band case, namely the eigenvectors of the Block states and the orbital character of the disorder. The latter can be described in general by promoting Eq. \eqref{Vi} to $V^{(mm')}_i=v_I \sum_{c}^{N_I} W_i^{mm'} \delta_{i j_c}$,  where $m,m'$ label orbital indices. We will assume as before that impurity centers are randomly located without correlation and that all impurities are of the same nature. We will explore below the consequences of different kind of disorder ensembles. Then, considering the self-average of disorder the square modulus in Eq. \pref{Qmulti1} evaluates as:
	\begin{widetext}
		\begin{eqnarray} \lb{multi_squaremod}
			\left\vert \langle {\bk},b \vert \sum_{i} {\vec c}_i^{\;\dagger} \cdot V_{i} \cdot\,{\vec c}_{i}  \;\vert \bk',b' \rangle \right\vert^2 &=&
			\left\vert \sum_{j}^{N_L} \sum_{mm'}^{N_b} U_{\bk}^{\dagger bm}U_{\bk'}^{ m'b'} v_{I} W^{mm'}_j \frac{e^{i\bx_j\cdot(\bk-\bk')}}{N_L}\;   \langle \bk,b \vert \,c^{b\,\dagger}_\bk\,c^{b'}_{\bk'}  \;\vert \bk',b' \rangle \right\vert^2 \nn\\
			&\simeq&  v^2_{I}\sum_{jl}\sum_{mm'nn'}\; \langle W^{mm'}_jW^{*nn'}_l\rangle_V \,\frac{e^{i(\bx_j-\bx_l)\cdot(\bk-\bk')}}{N_L^2}  \,P_{\bk,b}^{*mn}P_{\bk',b'}^{m'n'} \nn\\
			&=& N_I v^2_{I}\sum_{mm'nn'}\; D_{nn'}^{mm'} \;  \,P_{\bk,b}^{*mn}P_{\bk',b'}^{m'n'}
		\end{eqnarray}
	\end{widetext}
	In Eq.\ \pref{multi_squaremod} $U$ is the matrix which performs the rotation from the orbital to the band operators $c_\bk^b$, such that $c^m_j =\sum_\bk U_{\bk}^{ mb} c^b_\bk e^{i\bx_j\cdot\bk}/\sqrt{N_L}$. The $P$ are the corresponding  (Hermitian) projectors on the eigenvectors subspace $P_{\bk,b}^{mn} = U_{\bk}^{ mb}U_{\bk}^{\dagger bn}$, and we also introduced the dimensionless "disorder" tensor
	\be \label{dis1imp}
	D_{nn'}^{mm'} = W^{mm'}W^{*nn'}. 
	\ee
	
	Grouping the pair of indices $(mn)$ and $(m'n')$ as single indices, the summation in the last line of Eq. \ref{multi_squaremod} is equivalent to the vector-matrix-vector product. We will use the $\circ$ symbol to denote this product instead of the $\cdot$ symbol, to remind it is actually a tensor contraction. Then, the multiband collision integral kernel is concisely written as
	\be \lb{final_multi_Q}
	Q^{bb'}_{\bf k k'} = \kappa \,P^*_{\bk,b} \circ D \circ P_{\bk',b'} \; \delta(\xi_{{\bf k},b}-\xi_{{\bf k'},b'})
	\ee
	
	Notice that $D$ is always positive when fully contracted with projectors via the $\circ$ product. The quasiparticles rates are
	\begin{equation} \label{gamma_def}
		\G_{\bk,b} = \kappa  \, P^*_{\bk,b} \;\circ  \;D \;\circ \!\!\!\!\sum_{b',\bk'\in S(\varepsilon^{b'}_\bk)}  \!\!\!\! P_{\bk',b'}/|\bv_{\bk',b'}|
	\end{equation}
	where $S^b(\varepsilon)$ denotes the surface of energy $\varepsilon$ for the $b$-th band.
	 Here and later the notation $\sum_{b,\bk\in S^b(\varepsilon)}$ is a shorthand for $\mathcal{V}/(2\pi)^3\sum_{b}\int_{S^b(\varepsilon)}\mathrm{d}^2\bk$ which has the dimension of a length. Notice that we have used also the equivalences $\delta(\xi_{\bk,b}-\xi) = \delta(\e_{\bk,b}-\e) = \int_{S^b(\varepsilon)} \mathrm{d}^2\bk'\;\delta(\bk-\bk')/|v_{\bk',b}|$.
	Eq.\ \pref{gamma_def} shows the peculiar effect of the multiband structure: even for localized impurities, which imply an isotropic i.e. momentum-independent $D$ tensor (see Eq. \eqref{dis1imp}), the rotation 
	from the orbital to the band basis induces an effective momentum-dependence of the scattering kernel in the band basis, leading in general to both 
	a momentum-dependent scattering rate and, as we will see, a finite velocity renormalization, in contrast to the single-band case. Such a momentum dependence only 
	disappears when disorder tensor $D$ equals $\delta_{mn}\delta_{m'n'}$ so that $P^*_{\bk,b}\circ D\circ P_{\bk',b'}=Tr(P^*_{\bk,b})Tr(P_{\bk',b'})=1$ and one is left with the same structure \pref{defkappa} of the 
	single-band case. This condition is realized for GUE disorder, as we will discuss in Sec. \ref{disorder} below. 
	
	\subsection{Kernel inversion}
	In order to derive the dressed velocities $\bf{w}_\bk^b$ from Eq.\ \pref{w_solver} we need to invert the operator $(1-Q\tau)$. Before doing this, let us first observe that the kernel $Q\t$ has a number of properties. From the definition \pref{gkb} of the scattering rates it is evident that also in the more general case of inelastic scattering the kernel $Q\tau$ has unit eigenvalue, the right and left eigenfunctions being $w^{R}_{\bk,b} = 1/\t_{\bk,b}$ and $w^{L}_{\bk,b} = 1$ respectively. In the present case where scattering is elastic the unit eigenvalue is actually infinitely degenerate and the corresponding eigenfunctions can be labelled by the energy $\varepsilon$:
	\begin{eqnarray} \label{wrl}
	    w^{R}_{\bk,b,\varepsilon} &=& \delta(\varepsilon_{\bk,b} - \varepsilon) /\t_{\bk,b} \nn\\
	    w^{L}_{\bk,b,\varepsilon} &=& \delta(\varepsilon_{{\bf k},b}-\varepsilon).
	\end{eqnarray}
	Thus in analogy to solutions of inhomogeneous ODEs, for each velocity direction $\a$ the component $w^\a$ of $\bf w$ will be written as $(1-Q\t)^{-1} v^\a$ \textit{plus} a linear combination of the set $\{w^{R}_\varepsilon\}$. The first term, $(1-Q\t)^{-1}v^\a$, may in principle be divergent if a $v^\a$ is not orthogonal to $w^L_\varepsilon$ that is $\sum_{\bk,b} ({\bf v} \, w^L_\varepsilon)^b_\bk\neq0$, however one can check this is not the case and $\bf w$ is well-defined (for the proof of orthogonality see below Eq. \ref{vanishing}).
	
	For actual calculations the formal expression $(1-Q\tau)^{-1}$ needs to be evaluated. Since we are interested in an exact calculation one may na\"ively attempt to diagonalize the operator $Q\tau$. Unfortunately, this task cannot be achieved in a simple and direct way as for small matrices. Looking at the specific structure of the operator, we will rely on a mathematical trick to make progresses.
	Since the other eigenvalues of $Q\t$ have generically modulus less than unity \cite{Note2}, we can write the geometric series expansion $(1-Q\tau)^{-1}=\sum_{N\geq 0} (Q\tau)^N$. Writing one by one the terms we observe that the peculiar sandwich of the tensor $D$ with the matrices $P$ appearing in $Q$ (see Eq. \eqref{final_multi_Q}) allows for the definition of a tensor of size $N_b\times N_b\times N_b\times N_b $, we call it $K_\varepsilon$, which is a function of the energy only. Thus, we can rewrite $(1-Q\tau)^{-1}$ as:

	
	\begin{widetext}
		\begin{eqnarray} \lb{sol}
			(\delta_{\bf k\bf k'}\delta_{bb'} - Q^{bb'}_{\bf k\bf k'}\t^{b'}_{\bf k'})^{-1} &=& \sum_{N\geq0} ((Q\t)^N)^{bb'}_{\bf k\bf k'} =\delta_{\bf k\bf k'}\delta_{bb'} + P^*_{\bk,b} \circ D \circ \sum_{N\geq0} (K_{\varepsilon_\bk} \circ D)^N \circ 
			P_{\bk',b'}\tau^{b'}_{\bk'} \delta(\xi_{\bk,b}-\xi_{\bk',b'}) \nn\\
			&=& \delta_{\bf k\bf k'}\delta_{bb'} + P^*_{\bk,b} \circ D \circ
			\left(\mathbb{1} - K_{\varepsilon_\bk} \right)^{-1}
			\circ P_{\bk',b'}\,\tau^{b'}_{\bk'}\, \delta(\xi_{\bk,b}-\xi_{\bk',b'}) 
		\end{eqnarray}
	\end{widetext}
	where  $\mathbb{1} = \delta_{mm'}\delta_{nn'}$ and we defined the dimensionless tensor
	\begin{equation} \lb{Kmat}
		(K_\varepsilon)^{mm'}_{nn'} = \kappa \;\sum_{b,\bk\in S^b(\varepsilon)} 
		\left(P_{\bk,b} \right)^{mn}\,\frac{\t_{\bk,b}}{|\bv_{\bk,b}|} \,\left(P^*_{\bk,b} \circ D \right)^{m'n'}
	\end{equation}
	acting on the orbital indices.
	
	Roughly speaking what we did was to trade the inversion of an operator with the inversion of a tensor thus making an extremely convenient move. The relation between $Q\t$ and $K$ resembles the one between a matrix and its dual in the context of Random Matrix Theory\cite{For10}. For instance, dual matrices are employed when dealing with Wishart matrices and have been used to obtain the distribution of lifetimes of electrons in chaotic quantum dots with chiral symmetry\cite{Sch15} or to describe the topological physics of phonons in isostatic lattices\cite{Kane14}. When a positive matrix has the form $X^\dagger X$, with $X$ a rectangular matrix, its dual partner is the matrix $X X^\dagger$ which shares the same non-vanishing eigenvalues. 
	Here we have a similar structure. Despite the presence of the tensor $D$, which however determines only a small technical variation, the role of the $X^{a,b}$ matrix is played here by the projector $P^{(mn),({\bf k} b)}$. Moreover the eigenvalues and eigenvectors of $Q\t$ and those of $K_\varepsilon$ are in one-to-one correspondence, as shown in Sec. \eqref{App: TK1}.

	\subsection{Dressed velocities}
	We obtain the dressed velocities by combining expression \pref{sol} with Eq.\ \pref{w_solver}. Such velocities appear naturally as the sum of the bare ones, due to the first term in the second line of Eq. \eqref{sol}, a correction due to scattering, due the second term of the same equation, and a term due to inclusion of the "homogeneous" solutions $w^{R}_\e$ (as discussed below Eq. \eqref{wrl}):
	\begin{equation} \lb{multi_w}
		{\bf w}_{\bk,b}  =
		\bv_{\bk,b} + \kappa \; P^*_{\bk,b}  \circ D \circ \left(\mathbb{1} - K_{\varepsilon_\bk} \right)^{-1} \circ  {\bf F}_{\varepsilon_\bk} + \frac{1}{\t_{\bk,b}}\boldsymbol{\lambda}_{\varepsilon_\bk^b}
	\end{equation}
	Here $\bf F$ represents the on-shell average of the $P$ projector weighted with the bare velocity and the lifetime of the states as given by\cite{Note3}
	\begin{equation} \label{Fdef}
		{\bf F}_\varepsilon = \sum_{b',\bk\in S^{b}(\varepsilon)} 
		\left(P\,\tau\frac{\bv}{|\bv|}\right)_{\bk',b'}.
	\end{equation} 
	In the third term of Eq. \eqref{multi_w}, for each velocity direction $\a$, $\{\lambda_\varepsilon^\a\}$ is the set of linear combination coefficients for the set $\{w^{R}_\varepsilon\}$ shown in Eq. \eqref{wrl}. $\boldsymbol{\lambda}_\varepsilon$ is an on-shell-averaged vector with the dimension of a length and its precise value is determined in App. \ref{App:w0}. This last quantity ensures no variation of the total electrons density in each energy shell induced by the electric field. $\boldsymbol{\lambda}_\varepsilon$ vanishes identically in single band systems and, importantly, in time-reversal symmetric ones (see App. \ref{TRS}).

	Eq. \pref{multi_w} is the first major result of the paper. 
	A na\"ive strategy to solve Eq. \pref{w_solver} would have been to to discretize the Brillouin zone and diagonalize $Q\t$ or invert the matrix $1-Q\t$ in the grouped indexes $(\bk b)$ and $(\bk'b')$. This approaches would be computationally very demanding as the matrix size is $N_b N_k$, with $N_k$ the number of $k$-points. Roughly speaking, our approach allowed to fully solve the part involving momenta, while leaving behind only the inversion in the orbital space 
	of a matrix of "small" size $N_b^2$. 
	
	Since $F\sim \tau\sim 1/\kappa$ the second term in Eq.\ \pref{multi_w} is 
	of the same order as the first one in the disorder strength $\kappa$, therefore both contributions are equally relevant in most problems. However, 
	they manifest fundamentally different physics: the bare velocities depend 
	on band energies and their symmetry, while the corrections are controlled 
	by the electronic eigenvectors through $P_{\bk,b}$. 
	The condition by which the RTA is exact, i.e. $\bf w \parallel v$ and the transport is simply controlled by the lifetimes $\tau^{tr}$, does not yield a 
	physically transparent condition on $K$ and $F$, therefore we shall not show it. Notwithstanding, it is straightforward to see that the conjunction of certain symmetries (see App. \ref{App: Syms}), rotation and mirror ones for instance, might force such a collinearity of velocities as we will see in Sec. \ref{sec:Rashba} and even determine the vanishing of the vertex corrections ${\bf F}=0$.
	

	For the single band case the on-shell average of the dressed current ${\bf w}_\varepsilon := (1/\mathcal{N}(\varepsilon))\sum_{b,\bk\in S^b(\varepsilon)} ({\bf w}/|\bv|)_{\bk,b}$ vanishes. Using the fact that $\sum_{b,\bk\in S^b(\varepsilon)} (P/|\bv|)_{\bk,b}$ is the left-eigenmatrix of $K$ with vanishing eigenvalue, it is possible to generalize this sum rule. For the multiband case one has $ {\bf w}_\varepsilon = \Gamma_\varepsilon \boldsymbol{\l}_{\varepsilon}$ where we define the on-shell averaged scattering rate $ \Gamma_\varepsilon = (1/\mathcal{N}(\varepsilon)) \sum_{b,\bk\in S^b(\varepsilon)} (\G/|\bv|)_{\bk,b}$. Interestingly ${\bf w}_\varepsilon$ does not get contributions from neither the bare velocity nor the vertex correction terms. 
	We interpret such finding as the fact that direct scattering events (vertex corrections) preserve the symmetry of the original system where no preferential direction exists, whereas the charge redistribution implied by entropy maximization does not have such symmetry information.
	
	\subsection{Modelling the disorder: statistical ensembles for heterogeneous impurities}\label{disorder}
	
	The disorder tensor $D$ is highly material dependent. Moreover, most often impurities are not of the same kind thus several different matrices $W^{mm'}$ must be taken into account in the scattering. The self-averaging of the product $WW$ in Eq. \eqref{multi_squaremod} allows us to assume that each impurity $W^{mm'}$ is drawn from a statistical ensemble. This assumption makes sense also if one is not interested in the effect of a specific impurity type of disorder but rather in the effect of a generic class of disorder on a material. Without modifying the randomness in the location of the impurities, we assume that the matrix ensemble is Gaussian for each impurity, that is, we upgrade the average $\langle\rangle_V$ to
	\begin{eqnarray} \lb{Vprobs}
		\langle\cdot\rangle_V = \frac{1}{N_L^{N_I}}\prod_{c}^{N_I}\sum_{j_c}^{N_L}\;\langle\cdot\rangle_{W_c}
	\end{eqnarray}
	with
	\begin{eqnarray}  \lb{Vprobs2}
		\langle\cdot\rangle_W = \int {\mathcal D} \,W \;e^{ - Tr\,W^2/(2\sigma^2_W)}.\nn
	\end{eqnarray}
	
	We will assume that the matrix integration runs over one of the Wigner random-matrix ensembles, i.e. the Gaussian Unitary/Orthogonal/Symplectic Ensembles (GUE/GOE/GSE), or the Isotropic Intra-orbital Ensemble (IIE) whose precise definitions are confined to App. \ref{app: ensembles2}. The rationale behind the use of these ensembles is that they extend naturally the notion of statistical isotropy in real space to that in orbital space, as the eigenvectors of each $W_c$ are uniformly distributed in the corresponding projective spaces. For real materials no random sampling of the disorder configurations is expected to give exactly such ensembles, however they are suitable for calculations and help having physical insight on the effect on the combined action of symmetry and disorder (see App. \ref{app: ensembles2}).
	
	Adopting these ensembles, the disorder tensor must also be redefined as $D_{nn'}^{mm'}= \langle W^{mm'}W^{*nn'} \rangle_W$. For each ensemble the integral is easily computed (see App. \ref{app: ensembles3} for the derivation): 
	\begin{eqnarray} \label{Tstats}
		D^{mm'}_{nn'} \!\!&=&\! \sigma^2_W\begin{cases}
			\frac{1}{N_b}\delta_{mm'}\delta_{nn'} & \quad\quad\quad\mathrm{IIE}\\
			\delta_{mn}\delta_{m'n'} & \quad\quad\;\mathrm{GUE} \\
			\frac{1}{2}\left(\delta_{mn}\delta_{m'n'} + \delta_{mn'}\delta_{m'n}\right) & \quad\quad\;\mathrm{GOE} 
		\end{cases}\nn\\
		D^{(mq)(m'q')}_{(nr)(n'r')} \!\!&=&\!\!\frac{\sigma^2_W}{2}\bigg[\delta_{mn}\delta_{m'n'}\delta_{qr}\delta_{q'r'} \hspace{76pt}\mathrm{GSE}\nn\\
		&& + \delta_{mn'}\delta_{m'n}\delta_{q\bar r}\delta_{q'\bar r'}(-1)^{q+q'} \nn\\ &&+\delta_{mm'}\delta_{mn}\delta_{m'n'}\delta_{\bar qq'}(\delta_{q\bar r}\delta_{q'\bar r'} - \delta_{qr}\delta_{q'r'})\bigg] \nn\\
	\end{eqnarray}
	where the labels $q,q',r,r'=0,1$ refer to the spin-like degree of freedom (i.e. the quaternionic one) respectively attached to the $m,m',n,n'$ ones; we denoted $\bar x = 1-x$. 
	
	At first sight the "simplest" ensembles are the IIE and the GUE. In the first case $D$ acts as an identity w.r.t. (i.e. with respect to) the $\circ$ product and the elements of the collision kernel $Q^{bb'}_{\bf k k'}$ are simply the square of the overlap of scattered states. In the second case things are even simpler as these elements become functions of the energy shell of the scattered states only (and not of each momentum and band). Thus the GUE 
	behaves as a single-band result. In particular
	$\G^{b\,(GUE)}_{\bk} = \G^{(single)}_{\bk}$ and the velocity vertex corrections \textit{always} vanish, cf. with Eq. \eqref{G_single} and see App. \ref{App: TK2} for the proof.

	\section{dc conductivity tensor} \label{Sec: cond}
	Since there is no current flowing in the system at ${\bf E}=0$, at finite field only $\rho_{\bf E}$ contributes. This happens because $f_{\varepsilon}$ is a function of the energy only and the on-shell average of the band velocity vanishes. 
	For the same reason also the contribution from $\boldsymbol{\lambda}_\varepsilon$ vanishes. 
	Finally, for the current density, defined as $	{\bf J} = \frac{-e}{N_L a^3}\sum_{\substack{\bk,b}} \left(\bv\,\rho\right)_{\bk,b}$, we have 
	\begin{eqnarray*}
		{\bf J} &=& \frac{e^2}{N_L a^3} \sum_{\bk,b} \bv_{\bk,b}\,\tau_{\bk,b} \,({\bf w}_{\bk,b} \cdot {\bf E}) \left(-\partial_{\varepsilon^b_{\bk}} f_{\varepsilon^b_{\bf k}}\right)
	\end{eqnarray*}
	with $a$ the lattice constant.
	We can infer the conductivity tensor, defined as ${\bf J}=\s \cdot {\bf E}$, consisting of two different contributions:
	\begin{eqnarray} \label{sigma_cond}
		& & \quad\quad \sigma_{ij} =  \sigma^{bare}_{ij} + \sigma^{corr}_{ij} \\ \nn \\
		\sigma^{bare}_{ij} &=&  \frac{e^2}{N_L a^3}\int_\varepsilon \left(-\partial_{\varepsilon} f_{\varepsilon}\right) \sum_{b,\bk\in S^{b}(\varepsilon)}\left( v^i\,\frac{\tau}{|\bv|}\,v^j \right)_{\bk,b} \nn \\
		\sigma^{corr}_{ij}  &=& \frac{e^2\,\kappa }{N_L a^3} \int_\varepsilon  \left(-\partial_{\varepsilon} f_{\varepsilon}\right)
		\;F^{i*}_{\varepsilon}  \circ D \circ \left(\mathbb{1} - K_{\varepsilon} \right)^{-1} \circ  F^j_{\varepsilon} \nn
	\end{eqnarray}
	Here we label the two terms $\sigma^{bare}$ and $\sigma^{corr}$ for the analogy with the usual diagrammatic approach\cite{Mah00,Sch02,Bros16}. The first term is the sum of the conductivities produced by each single electronic state, equipped only with its finite lifetime due to disorder. In the diagrammatic language it corresponds to the bubble with insertion of bare velocity operators at each vertex, and then it is called the 
{\it bare bubble} term.  The second term is \textit{the} peculiar feature of the multiband models, and it is equivalent in the diagrammatic language to a {\it correction} due to renormalization of the velocity vertex. It corresponds to a bubble containing a bare velocity vertex and a dressed velocity vertex, which originates from the so-called vertex corrections\cite{Mah00}. It comes from the disorder-mediated overlaps between the pairs of electronic eigenstates produced by the electric field. As opposed to the first term it does not really contain single-particle contributions but rather two-particles ones. Thus, we stress that it is generically misleading to interpret the linear response of 
	a multiband system as the mere sum of the independent contributions of the quasiparticles or of the bands.
	
	It is a remarkable fact that the conductivity gets \textit{always} enhanced by the vertex corrections, that is $\sigma^{corr}\geq 0$, whenever the 
	disorder tensor $D$ is $\circ$-positive (see App. \ref{App: TK3} for the proof). For instance, this is true with the GUE or any kind of purely intra-orbital disorder, like the IIE one. In general no weak- or weak anti-localization effects\cite{Han18}, let alone Anderson localization\cite{Mir08}, are expected to be seen within this model as they emerge only when coherent multiple scattering is taken into account.
	
	Using the property $A \circ X \circ B =B\circ X^* \circ A$, with $X=D,K$ and $A,B$ any two matrices, and the reality of integrand in $\sigma^{corr}$ it is easy to show that the Onsager relation\cite{Nag10} $\sigma_{ij} =  \sigma_{ji}$ is satisfied always irrespectively of whether the system is time-reversal symmetric or not. The strange robustness of this relation comes from the exclusion of magnetic field driving terms in our Boltzmann equation; the terms coming from its explicit inclusion are expected to lead to the more general formula $\sigma_{ij}({\bf B}) =  \sigma_{ji}(-{\bf B})$. 
	
	\section{Disorder engineering} \label{sec:dis_eng}
	We have answered so far the question: What are the transport properties of a system given a specific kind of disorder? We can as well ask ourselves the converse\cite{Ash02,Bra06}: to what extent can disorder affect transport properties? Can we tailor specific desired features? First we focus our attention on the rates. We claim that i) it exists always a disorder that makes rates constant at a specific energy shell; ii) it is always possible to set to zero the rates at $N_b-1$ distinct points of the Fermi surface. Notice that the second statement may be used to try to engineer a dc conductance which is anisotropic, for instance by setting to zero the rates at those momenta at which velocities in one direction are the highest.
	Statements about vertex corrections, instead, are more difficult to establish. The only general and non-trivial statement that can be easily done is that for a generic non-symmetric system if $D$ has no vanishing $\circ$-eigenvalues then it is impossible to make vertex corrections to vanish, even for a single direction. All proofs are confined in App. \ref{App:dis_eng_proofs}.

	\section{Example: 2D Rashba electron gas} \label{sec:Rashba}
	
	\begin{figure}[!]
		\includegraphics[width=0.45\textwidth]{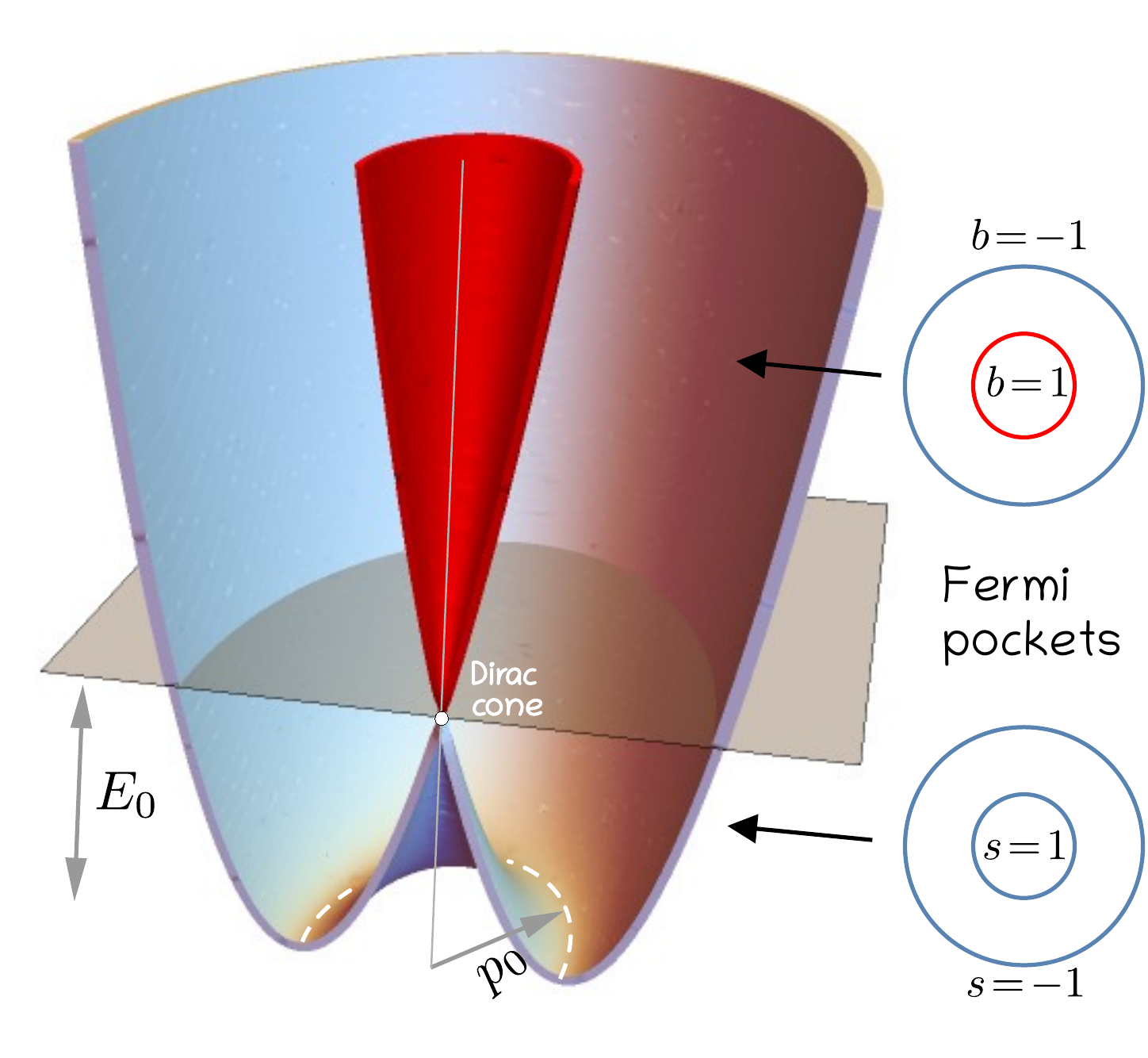}
		\caption{Dispersion of a 2D Rasbha electron gas featuring a Dirac cone whose energy level $E_0$ is marked by a gray plane. On the right, a sketch of the Fermi pockets in the Brillouin zone is shown. The label $b$ refers to the bands (coloured in red and light blue) and is used in the high-density regime while the label $s$ is used to distinguish the two pockets in the low-density regime.} \label{fig:Rashba_disp}
	\end{figure}
	
	As a specific example of multiband system we consider a 2D electron gas with intrinsic spin-orbit coupling ruled by the so-called Rashba Hamiltonian\cite{Byc84,Sch02,Rai12,Bros16,Cong16}. This system is quite simple in itself; however, it allows us to discuss in a full analytical fashion all constituents of the dressed velocity and the conductivity formulas. The problem was partly addressed in Ref. \onlinecite{Bros16}, where multiorbital effects within both a quantum and Boltzmann approach were discussed for the IIE type of disorder. Here we provide a full  analysis of the interplay between different disorder realizations, symmetry and the two-orbital character of the Fermi sheets, and we discuss how transport properties can be used to gain insight into the nature of spin-orbit interactions. 
	A more complicated and less symmetric system will be presented somewhere else\cite{Note1}.
	
	Let's consider a 2D system of area $\mathcal V$. The Rashba Hamiltonian in the vicinity of $\bk=0$ is
	\begin{equation}   \label{Hrashba}
	H(\bk) = \frac{k^2}{2m} + \alpha \hat z \cdot \bk \times \vec \sigma + E_0 	    
	\end{equation}
	where $\sigma$ acts on a spin 1/2 degree of freedom and $E_0 = p_0^2/(2m) = m \alpha^2/2$ is an energy offset.
	The Hamiltonian can be diagonalized as $ U^{l} \Lambda_l \,U^{l\,\dagger}$ with (we use polar coordinates for $\bk$):
	\begin{eqnarray} \label{rashba_diag}
		U(\theta) \!=\!  \frac{1}{\sqrt{2}}\left(
		\begin{array}{cc}
			e^{i\theta}& e^{i\theta}   \\
			i  & -i  \\
		\end{array}
		\right)\!,\,\Lambda\! =\! \mathrm{diag}\left(\frac{(k+p_0)^2}{2m} ,\;\frac{(k-p_0)^2}{2m}  \right) \nn\\
	\end{eqnarray}
	
	The system has time-reversal symmetry, continuous rotation symmetry and mirror symmetries about both the x and y axes (see Sec. \ref{App: Syms}). 
	The two bands are shown in Fig. \ref{fig:Rashba_disp}. A Dirac cone is present in the middle at energy $E_0$ where the two bands touch. At each energy the full Fermi surface is made of two concentric circles, however, we distinguish two transport regimes characterized by the values of the chemical potential $\mu>E_0$ (high-density regime) and $\mu<E_0$ (low-density regime). The eigenvectors of $H$ (i.e. the columns of $U$) both transform as $U_{\cdot \,i}(\theta) = \sum_j U^R_{\cdot\, j}  U_{j i}(0) \quad (i=1,2)$ with $U^R$ a representation of the 2D rotation in $\mathbb{C}^2$: 
	\[U^R= \begin{pmatrix}
		e^{-i\theta} & 0 \\
		0 &        1
	\end{pmatrix},
	\]
	or $U_{\cdot \,i}(\theta) \propto \sum_j U^{M_x}_{\cdot\, j}  U_{j i}(-\theta) \propto \sum_j U^{M_y}_{\cdot\, j}  U_{j i}(\pi-\theta),  \quad (i=1,2)$ and with $U^{M_{x,y}} = \sigma_{y,x}$ a representation of the mirror symmetries. 
    Concerning disorder, we consider the IIE, the GUE and (since the system is spinfull) the GSE, that we call in the following RMT-disordered systems (where RMT stands for Random-Matrix Theory). In addition we also explore the effect of a disorder characterized by a Single-matrix Impurity Disorder (SID) $W$, that we {\it engineer} in order to break all symmetries the Hamiltonian Eq. \eqref{Hrashba}. For instance, we will consider
	\[
	W =  \sqrt{\frac{2}{3}}\left(
		\begin{array}{cc}
			0 & 1  \\
			1 & 2 \\
		\end{array}
		\right).
	\] 
	Practically, such a disorder may be caused by a specific kind of magnetic impurity, able to break not only the time reversal symmetry but also the $x/y$ isotropy. 
	Before proceeding we notice that the GSE coincides with the IIE in this case where $N_b=2$ and we will call both under the label of the latter ensemble in the following. This coincidence happens because the Kramers degeneracy\cite{For10,Meh04} in the GSE imposes vanishing disorder-induced hopping between different spin species and forces the diagonal elements of the disorder to be equal, thus making the matrix proportional to the identity (thus indistinguishable from IIE matrices). In the following, we will represent tensors as matrices according to the mapping $T^{m+N_b n\,,\,m'+N_b n'} = T^{m,m'}_{n,n'}$, where for the sake of brevity we assumed the numbering of the labels are from $0$ and $N_b-1$. In particular the disorder matrices derived from Eq. \eqref{Tstats} and Eq. \eqref{dis1imp} for the are:
	\begin{eqnarray}
		D^{(GUE)} &=& \left(
		\begin{array}{cccc}
			1 & 0 & 0 & 1 \\
			0 & 0 & 0 & 0 \\
			0 & 0 & 0 & 0 \\
			1 & 0 & 0 & 1 \\
		\end{array}
		\right), \quad D^{(IIE)} = \mathrm{diag}(2,2,2,2) \nn \\
		D^{(SID)} &=&\frac{2}{3}\left(
    \begin{array}{cccc}
     0 & 0 & 0 & 1 \\
     0 & 0 & 1 & 2 \\
     0 & 1 & 0 & 2 \\
     1 & 2 & 2 & 4 \\
    \end{array}
    \right) \nn
	\end{eqnarray}
	Here we choose to work with the set of $\s_W$'s in \eqref{App:choice2}; in the SID case the factor $2/3$ is there to guarantee indeed an average disorder strength equal to that of the other ensembles, in the spirit of the chosen normalization, see Sec. \ref{App:ensComparison}. 
	Notice that the matrix is semi-positive definite for the RMT-disordered systems, that will guarantee a semi-positive definite correction to the conductance, see Sec. \ref{Sec: cond}. Moreover, these matrices obey the symmetry of the clean system while for the SID all symmetries are broken, see Eqs. \eqref{App:dis_sym} and \eqref{Trev_invD}. Thus, the RMT-disordered systems, having the full system mirror and a rotation symmetry, must lead to dressed velocities parallel to the bare ones and we expect the RTA to be exact. 
	
	We are not interested here in the temperature effects, so we work at $T=0$.
	
	\subsection{High-density regime}
	We focus first on the case with $\mu>E_0$. The Fermi circles pertain to the two different bands, we label them with $b=\pm1$ (accordingly to the sign of $k\pm p_0$ in $\Lambda$ in Eq.\eqref{rashba_diag}, see also Fig. \ref{fig:Rashba_disp}). We have
	\begin{eqnarray}
		\bv_{b,\theta} &=& v_F (\cos\theta,\sin\theta),\quad
		v_F = \sqrt{2\mu/m} \nonumber \\
		k_{F b} &=&  m\,v_F -b\,p_0. \nn
	\end{eqnarray}
	here $k_{Fb}$ denotes each Fermi momentum for each circle.
	Every summation over isoenergetic momenta $\sum_{b,\bk\in S^b(\varepsilon)}$ may be replaced by  $\mathcal V \sum_b  \int_0^{2\pi} \mathrm d \theta \frac{k_{Fb}}{(2\pi)^2}$. 
	The projectors obtained by Eq. \eqref{rashba_diag} are
 	\begin{eqnarray}
		P_{b,\theta}  \!\!&=&\!\!  
		\frac{1}{2}\left(\sigma_0 + b\cos\theta\sigma_y +b\sin\theta\sigma_x\right)
		= 	\frac{1}{2} \left(
		\begin{array}{cc}
			1 &  -i\,be^{i\theta}  \\
			i\,b e^{-i\theta} & 1  \\
		\end{array}\right) \nn\\
		&\rightarrow& 
		\vec P_{b,\theta}  = 
		\frac{1}{2} \left(
		\begin{array}{c}
			1 \\
			-i\,b e^{i\theta}  \\
			i\,b e^{-i\theta} \\
			1  \\
		\end{array}
		\right) \nn
	\end{eqnarray}
	In the last step we mapped $P$ to a vector $\vec P^{j+N_b i} = P^{ij}$ (in analogy with the tensor mapping showed before).
	From Eq. \eqref{gamma_def} we compute momenta-independent rates
	\begin{eqnarray} \label{rates_HD}
	 \Gamma^{(GUE)}_{b,\theta} &=& \Gamma^{(IIE)}_{b,\theta} = \frac{\kappa m}{\pi}\nn\\
	 \Gamma^{(SID)}_{b,\theta} &=&\frac{\kappa m}{\pi}\left(1+\frac{2}{3} b \sin (\theta)\right)
	\end{eqnarray}
	
	The angle-averaged rate is the same for all kind of disorder types considered. This feature has to be connected with the $\s_W$-normalization we have chosen. This choice enforces same strength of the overall disorder rather than same strength per disorder degree of freedom (normalization \eqref{App:choice1}) which e.g. would produce  $\Gamma^{(GUE)}>\Gamma^{(IIE)}$ instead. Moreover, notice that $\Gamma^{(SID)}$ is band dependent, anisotropic and breaks all symmetries but the mirror symmetry $M_y$. The survival of this symmetry is accidental in the sense that is due to the specific angle dependence of the projectors and not to their global symmetry.
	Further, from Eq. \eqref{Fdef} we find
	\begin{eqnarray*}
		{\bf  F}^{(l)}_\varepsilon &=&  -\frac{\tau^{(l)}\,p_0}{4\pi} (\sigma_y,\sigma_x)  \nn\\
		\longrightarrow
		\vec F^{(l)}_x &=& \frac{\tau^{(l)}\,p_0}{4\pi} \begin{pmatrix}
			0  \\
			i \\ 
			-i  \\
			0 \\ 
		\end{pmatrix}, 
		\vec F^{(l)}_y =\frac{\tau^{(l)}\,p_0}{4\pi} \begin{pmatrix}
			0  \\
			-1 \\ 
			-1  \\
			0 \\ 
		\end{pmatrix}, \nn\\ 
		\vec F^{(SID)}_x &=&  \frac{3\left(3-\sqrt{5}\right)\,\tau^{(GUE)}p_0}{8\pi}\left(
\begin{array}{c}
 0 \\
 i \\
 -i \\
 0 \\
\end{array}
\right),\nn\\
\vec F^{(SID)}_y &=& 
\frac{3\sqrt{5}(3-\sqrt{5})\,\tau^{(GUE)} p_0}{20\pi}\left(\begin{array}{c}
 1 \\
 -3/2 \\
 -3/2 \\
 1 \\
\end{array}
\right)
	\end{eqnarray*} 
	where $l$ runs over the ensembles labels. We find as well 
	\begin{eqnarray*} 
		K^{(GUE)} &=&  \frac{1}{2}\left( 
		\begin{array}{cccc}
			1 &  0 & 0 & 1  \\
			0& 0 & 0 & 0  \\
			0& 0 & 0 & 0  \\
			1 &  0 & 0 & 1  \\
		\end{array}
		\right)  \nn\\
		K^{(IIE,GSE)} &=&  \frac{1}{2} \left( 
		\begin{array}{cccc}
			1 &  0 & 0 & 1  \\
			0& 1 & 0 & 0  \\
			0& 0 & 1 & 0  \\
			1 &  0 & 0 & 1  \\
		\end{array}
		\right) \nn\\
K^{(SID)} = 
\frac{1}{20}&&\!\!\!\!\!\!\!\left(
\begin{array}{cccc}
 2 \sqrt{5} & 5+\sqrt{5} & 5+\sqrt{5} & 20-\frac{10}{\sqrt{5}} \\
 5-3 \sqrt{5} & \sqrt{5}-5 & 10-\frac{20}{\sqrt{5}} & 3 \sqrt{5}-5 \\
 5-3 \sqrt{5} & 10-\frac{20}{\sqrt{5}} & \sqrt{5}-5 & 3 \sqrt{5}-5 \\
 2 \sqrt{5} & 5+\sqrt{5} & 5+\sqrt{5} & 20-\frac{10}{\sqrt{5}} \\
\end{array}
\right)
	\end{eqnarray*} 
	Due to the simplicity of the system, in particular the independence of the eigenstates from the energy, $K$ does not depend on energy as well.
	Notice that for the RMT-disordered systems, the $(1,4)$ and the $(2,3)$ blocks are decoupled both in $D$ and $K$. In particular, $K$ has a unit eigenvalue (as expected) coming from the first block. Importantly $F$ does not have any component on the $(1,4)$ sector and the renormalized velocities are readily found by inversion of the $(2,3)$ block alone. More difficult is the inversion in the SID case but still treatable analytically. One gets (vertex corrections are the second terms in the expression):
	\begin{eqnarray} \label{w_HD}
		{\bf w}_{b,\theta}^{(GUE)} &=&  \bv_{b,\theta} \nn\\
		{\bf w}_{b,\theta}^{(IIE)} &=& 	\bv_{b,\theta}-b\frac{\,p_0}{m v_F} \bv_{b,\theta} \nn\\	 
		{\bf w}_{b,\theta}^{(SID)} &=& \bv_{b,\theta} + b\frac{p_0}{mv_F}\left(\frac{\left(4-\sqrt{5}\right)}{11}v_{b,\theta}^x, \frac{\left(\sqrt{5}-2\right)}{3}v_{b,\theta}^y\right) \nn\\
	\end{eqnarray}
	Since for the RMT-disordered systems the renormalized velocities are collinear with the band velocities the RTA is exact in this regime. In the SID case the collinearity is lost, but the dressed velocity retains time-reversal and mirror symmetries. 
	The conductivity matrix is diagonal (in the RMT-disordered case with equal diagonal elements):
	\begin{eqnarray}  \label{rashba_cond_HD}
		\sigma^{(GUE)} &=&  \frac{e^2}{\kappa m}\mu\nn\\
		\sigma^{(IIE)} &=&  \frac{e^2}{\kappa m}[\mu + E_0] \nn\\	
		\sigma^{(SID)}_{xx} &=&  \frac{e^2}{m \kappa} \frac{3}{22} \left[11 \left(3-\sqrt{5}\right) \mu +\left(7 \sqrt{5}-17\right)E_0\right] \nn\\
		\sigma^{(SID)}_{yy} &=&
		\frac{e^2}{m\kappa}\frac{9}{10} \left[\left(3 \sqrt{5}-5\right) \mu+\left(33 \sqrt{5}-75\right)E_0 \right] \nn\\
	\end{eqnarray}
	where the first terms in the square brackets are the bare bubble contributions while the second ones (absent for the GUE) come from the current dressing. Notice how the conductivity corrections in the SID case are negative, which is compatible with the fact that $D$ is not $\circ$-positive definite.  
	
	\subsection{Low-density regime}
	In the case $\mu<E_0$ there are two Fermi pockets both pertaining to the second band, we will label them with $s=\pm1$ (using the plus sign for the inner circle, see Fig. \ref{fig:Rashba_disp}). We have
	\begin{eqnarray}
		v_s &=& -s\, v_F (\cos\theta,\sin\theta),\quad
		v_F = \sqrt{2\mu/m} \nn \\
		k_{F s} &=& \,p_0 - s \,m\,v_F  \nn
	\end{eqnarray}
	Again every summation over isoenergetic momenta $\sum_{s,\bk\in S^s(\varepsilon)}$ may be replaced by  $\mathcal V \sum_s  \int_0^{2\pi} \mathrm d \theta \frac{k_{Fs}}{(2\pi)^2}$. 
	The projectors are only angle dependent: $P_{s,\theta} = \frac{1}{2}\left(\sigma_0 -\cos\theta\sigma_y -\sin\theta\sigma_x\right)$.
	In contrast to the other regime, rates are energy dependent and diverge at the band bottom, where $v_F \rightarrow 0$: 
\begin{eqnarray} \label{rates_LD}
	 \Gamma^{(GUE)}_{b,\theta} &=& \Gamma^{(IIE)}_{b,\theta} = \frac{\kappa p_0}{\pi v_F}\nn\\
	 \Gamma^{(SID)}_{b,\theta} &=&\Gamma^{(GUE)}_{b,\theta}\left(1+\frac{2}{3} b \sin (\theta)\right)
	\end{eqnarray}
	Concerning $\bf F$, $K$ and $\bf w$, we can simply make the exchange $p_0 \leftrightarrow m v_F$ and the change $b \rightarrow s$. Notice that surprisingly $K$ stays the same in the two regimes. The conductivities are
	\begin{eqnarray} \label{rashba_cond_LD}
		\sigma^{(GUE)} &=&  \frac{e^2}{\kappa m}\mu\nn\\
		\sigma^{(IIE)} &=& \frac{e^2}{\kappa m}(\mu+ \frac{\mu^2}{E_0})\nn\\
        \sigma^{(SID)}_{xx} &=&  \frac{e^2}{m \kappa} \frac{3}{22} \left[11 \left(3-\sqrt{5}\right) \mu +\left(7 \sqrt{5}-17\right)\frac{\mu^2}{E_0}\right] \nn\\
		\sigma^{(SID)}_{yy} &=&
		\frac{e^2}{m\kappa}\frac{9}{10} \left[\left(3 \sqrt{5}-5\right) \mu+\left(33 \sqrt{5}-75\right)\frac{\mu^2}{E_0} \right] \nn\\
	\end{eqnarray}

	In analogy with the other regime, we can easily distinguish the bare contribution (first term) and the vertex correction one (second term).

	\begin{figure}[!]
		\includegraphics[width=0.45\textwidth]{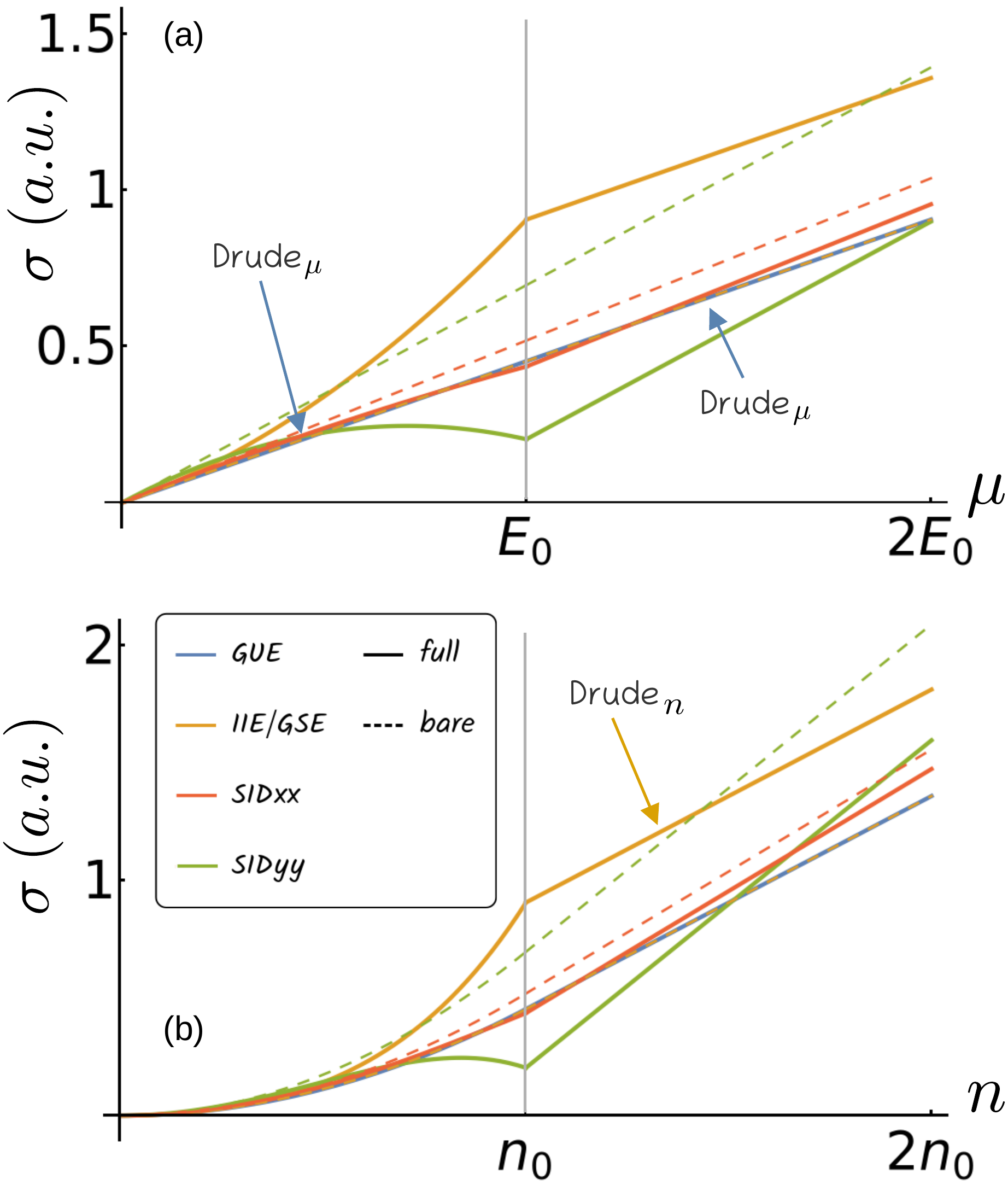}
		\caption{Dc conductivity of a Rashba metal as a function of the chemical potential (a) and the electronic density (b). The low-density regime lies below the values $E_0$ and $n_0$ respectively (grey lines). Different kinds of disorder are considered and both the bare and full conductivities are shown. Being the $\s^{SID}$ anisotropic, both xx and yy results are shown. The Drude regimes according to formulas Eqs. \eqref{Drude_for} and  \eqref{Drude_for2} are indicated respectively with the labels $\mathrm{Drude}_{n/\mu}$.} \label{fig:Rashba_cond}
	\end{figure}

	\subsection{Discussion}  \label{sec:RashbaC}
	
	
	Full and bare conductivities in Eqs. \eqref{rashba_cond_HD}  and \eqref{rashba_cond_LD} as a function of the chemical potential are shown in Fig. \ref{fig:Rashba_cond}(a) for the different ensembles.
	The functional dependence on $\mu$ and $E_0$ has the form $c_1 \mu+c_2E_0$ in the  high-density regime and $c_1 \mu+c_2 \mu^2/E_0$ in the low-density regime for all disorder types, the difference among them being only in the values of the prefactors $c_1,c_2$.  In addition notice that for a given disorder type these  prefactors stay the  same in the two regimes. We first observe that the magnitude of the slopes of the different curves is similar in the range shown. This is a consequence of the normalization Eq. \eqref{App:choice2} for $\s_W$. This guarantees the same norm for the tensor $D$ thus favoring similar rates, as already discussed, and similar conductivities. Different normalizations choices would have left the shapes unchanged, since $\s_W$ enters in $\sigma$ only as an overall prefactor (see Eqs. \eqref{sigma_cond} and \eqref{Tstats}), leading instead to different slopes.
	It is hard to predict which ensemble has the highest conductivity at fixed $\mu$. However, at high energy, where only the bare bubble term matters, the SID conductivities (both the one along $x$ and $y$) are the highest, due their higher average lifetimes as compared with those of the other ensembles (n.b. the average rates are instead the same).
	
	The contributions of the vertex correction terms can be appreciated from the figure as the difference between the continuous and the dashed lines. The corrections vanish for the GUE, are big and positive for the IIE and GSE and negative for the SID ensemble.
	The conductivities as a function of the electron density are shown in Fig. \ref{fig:Rashba_cond}(b), where one keeps in mind the (ensemble-independent) relations
	\begin{equation}\label{densities}
	n=\begin{cases} 
	    m (\mu + E_0)/\pi, &\mathrm{high\:density} \\
	    2m\sqrt{\mu E_0}/\pi,& \mathrm{low\:density}.
	\end{cases} 
	\end{equation}
	In the plot we have defined the density at $E_0$ as $n_0=2mE_0/\pi$.
	The IIE result has been derived previously in Ref. \onlinecite{Bros16}. There it was noticed the remarkable fact that the high-density conductivity coincides with the Drude one for a standard (i.e. non-Rashba) metallic parabolic band 
	\begin{align}  \lb{Drude_for}
	\s = e^2n\tau/m.
	\end{align}
	where for the IIE we can substitute $\tau \rightarrow \tau^{IIE}=1/\G^{IIE}\equiv \pi/(\kappa m)$, see Eq. \eqref{rates_HD}.
	Moreover, and by contrast, the same coincidence is not there in the low-density regime where the dependence in $n$ of the conductivity is nonlinear. Such change of the conductivity from following the Drude law to following an "unconventional" one was ascribed to the peculiar structure of the Fermi surface in the low-density regime. Indeed, since the scattering must preserve the spin some scattering processes are suppressed by the spin-momentum locking of the Rashba eigenstates.  Thus, the change of helicity of the inner pocket at $\mu<E_0$ explains the anomalous behavior of the conductivity in the low-density regime\cite{Bros16}. 
	
	The authors pointed out also the interesting fact that, in determining the Drude regime at high densities, the inclusion of the velocity corrections (encoded in the quantity $1-\tau/\tau^{tr}$ in their RTA language) was crucial, as the bare conductivity would lead again to an unconventional behavior (a "shifted" Drude law) $\s^{bare} = e^2(n-n_0/2)\tau^{IIE}/m$. 
	
	Applying the same analysis to the other ensembles, it is straightforward to see that the low-density regime still features unconventional transport. However, differently from the IIE case, the transport stays unconventional also at high density, since the prefactors of $\mu$ and $E_0$ in Eqs. \eqref{rashba_cond_HD}  and \eqref{rashba_cond_LD} are different (cf. Eq. \eqref{densities} top row). It is clear that also the bare conductivities do not follow the Drude behavior, since the absence of the term proportional to $E_0$ does not allow to reconstruct the value of $n$ as given by Eq.\ \eqref{densities}.
	
	To delve deeper into this matter about the relation between the Rashba and the Drude conductivity, and its possible observation in experiments, it is worth considering not only the case when the conductivity is measured against the electronic density (via for instance electronic doping) but also the case where it is measured with respect to the chemical potential (for instance, tuning a gate voltage in mesoscopic apparatuses). In this case we should compare our results with the expected Drude formula for a standard spinfull 2D metal  as a function of the chemical potential, i.e. 
	\begin{equation} \label{Drude_for2}
	\sigma = e^2\mu\tau/(2\pi),    
	\end{equation}
so that the conductivity is simply proportional to the chemical potential.
	As can be easily seen from Eqs. \eqref{rashba_cond_HD} and \eqref{rashba_cond_LD}, all ensembles save the GUE feature unconventional behavior (i.e. different from that of Eq. \eqref{Drude_for2}) at both low- and high-density regimes. 
	The GUE conductivity surprisingly follows the Drude formula \eqref{Drude_for2} in both regimes, assuming that we take for both regimes the same lifetime i.e. $\tau\rightarrow\tau^{GUE}=\pi/(\kappa m)$ (as defined in Eq. \eqref{rates_HD}).
	So, with GUE disorder the Rashba physics stays somehow "hidden" from a conductivity measurement varying the chemical potential: nothing special happens at the Dirac cone ($\mu=E_0$) and, more generally, the spin-orbit coupling does not affect any quantity involved in the conductivity formula, as if the system had $\alpha=0$. We may understand such phenomenon as follow. The GUE disorder fully suppresses the role of the electronic eigenstates at the Fermi level thus removing any scattering bias due to spin-momentum locking of helicity eigenstates and reducing the transport to be dependent only on the geometric shape of the dispersion as for the single-band case. It is not difficult to show that if a GUE-disordered system has rotational symmetry then $\sigma = \alpha e^2 v_F^2/\kappa$ with $\alpha$ some real number. The peculiarity of the Rashba model enters only in  the fact that the relation between the chemical potential and the group velocity is the same found in a standard metal i.e. $\mu=m v_F^2/2$. So, after defining $\tau=\alpha\pi/(\kappa m)$, one recovers the Drude law Eq. \eqref{Drude_for2}. 
	If one uses instead the density $n$ in the Drude formula rather than $\mu$, a departure of the GUE conductivity from Eq. \eqref{Drude_for} is found as $n$ keeps track of the shape of the bands dispersion (through Luttinger's theorem) which is clearly different in a standard and a Rashba metal.

	\section{conclusions and perspective}
	
	To summarise, we have solved the Boltzmann transport problem at small electric field and weak impurity-only scattering for a generic multiband system. In doing so we went beyond the formal solution involving the inverse of the collision-integral operator and showed that an analytical solution exists in terms of the inverse of a tensor whose dimension is set by the number of bands. Such a solution is naturally expressed as a correction to the bare band velocity, in full analogy with the  vertex-corrections term which appears in the diagrammatic approach. The resulting corrections to both the velocity and the dc conductance matrix have the same order of magnitude as the bare ones. 
	
	Much attention has been devoted in treating disorder. Its effect appears through a tensor that modifies the overlap between the scattering eigenstates unless it stems from a IIE where it simply sets the scattering magnitude. For such convenience this ensemble is the most employed one\cite{Rai12,Bros16,Cong16,Nag20}. As a specific example we analyzed the case of a 2D Rashba electron gas, where we also discussed the topic of "disorder engineering" discovering how a single-impurity disorder may be tailored to uniform the rates across the Fermi surfaces at specific energies. We also found that,  quite surprisingly, the conductivity of GUE-disordered model does not manifest the underlying Rashba physics if it is measured as a function of a gate voltage instead of the electronic density. 
	
	In perspective, it would be interesting to include in the model scattering from particle interactions, at least at RTA level. Such a generalization would open up the possibility to study at dept the interplay between disorder and interaction. For instance in a recent work\cite{Nag20} the temperature dependence of the resistivity of the Rashba gas is shown to have a different behavior above and below the Dirac point. Thus a relevant question is whether such property is robust under a different disorder ensemble in analogy to what we have discussed in the case without interactions. 
	
	Finally, other possible directions involve the finite-frequency problem, to make contact with spectroscopic observables, the generalization to nonlinear regime of the electric field, though diabatic effects will not be included in such a semiclassical theory, and the inclusion of magnetic field, for the study of the Hall effect and magnetoresistance. The hope is to provide accessible and analytical semiclassical formulas to facilitate the understanding of more advanced but often opaque and numerically demanding fully-quantum approaches.

	\section{acknowledgements}
This work has been supported by PRIN 2017 No. 2017Z8TS5B, by Sapienza University via Grant No. RM11916B56802AFE and RM120172A8CC7CC7, and by Regione Lazio
(L.R. 13/08) under project SIMAP.

	\begin{appendix} 
		
		\section{Determination of $\boldsymbol{\l}_\varepsilon$} \label{App:w0}
		
		To express the population correction Eq. \eqref{pop_corr} it is necessary to fix the quantity $\boldsymbol{\l}_\varepsilon$ in Eq. \eqref{multi_w} at each energy. In order to do that we invoke the maximum entropy principle subject to the total-energy and particle-number constraints in the following way. The functional to maximize is
		\begin{eqnarray}
			\mathcal{\tilde F}[	\rho_{\bk,b},\alpha,\beta] &=& -\sum_{\bk,b} \left[\rho_{\bk,b} \ln (\rho_{\bk,b}) +\left(1-\rho_{\bk,b} \right)\ln \left(1-\rho_{\bk,b} \right)\right] \nn \\
	& & + \alpha\left(\sum_{\bk,b} \rho_{\bk,b} - n^{el} \right) \nn \\
	& & + \delta\left({\bf E}\right)\beta \left(\sum_{\bk,b} \varepsilon_{\bk,b} \rho_{\bk,b} - h \right) \nn\\
		\end{eqnarray}
		where $n^{el}$ is the electron density which is left invariant by the action of the electric field otherwise charge would pile up accross the system violating the initial assumptions over the electric field to be small and constant. The same reasoning does not apply to the energy density $h$ which may change at finite field from its initial value. $\alpha$ and $\beta$ are two lagrangian multipliers. 
		
		Clearly at zero field the maximization yelds $\rho_\varepsilon = f_{\varepsilon}^{(\beta,\mu)}$, with $f$ the Fermi distribution at a certain temperature $\beta$ and chemical potential $\mu=\alpha/\beta$.
		
		At finite field the solution of the Boltzmann equation is imposed by demanding that the actual functional to maximize is
		\[
		\mathcal{F}[\boldsymbol{\l}_\varepsilon,\alpha,\beta] = \mathcal{\tilde F}[f_{\varepsilon} +  {\bf E}  \cdot \left(\boldsymbol{\l}_\varepsilon + \boldsymbol{\l}_{\bk,b}\right),\alpha,\beta]
		\]
		where $\boldsymbol{\l}_{\bk,b} = {\bf w}_{\bk,b} \,\t_{\bk,b} \,\partial_{\varepsilon^b_{\bf k}} f_{\varepsilon^b_{\bf k}}$, including in ${\bf w}_{\bk,b}$ only the first two terms in Eq. \eqref{multi_w}. Maximizing $\mathcal{F}$ and considering the linear order in $\bf E$, we get the system of equations
		\begin{eqnarray}&
			\begin{cases}
				\partial_{\rho_\varepsilon} \mathcal F = 0 \\
				\partial_{\alpha} \mathcal F = 0 \\
			\end{cases} \implies \nn\\
			&\begin{cases}			
				\boldsymbol{\l}_\varepsilon = -\left(\nabla_{\bf E}\alpha\right)/\left[2\mathcal{N}(\varepsilon)\left(1+\cosh(\beta(\varepsilon-\mu))\right)\right] -\\
				\quad\quad\;\:\,- \sum_{b,\bk\in S^{b}(\varepsilon)} \boldsymbol{\l}_{\bk,b} /\mathcal{N}(\varepsilon) \\
				\int_\varepsilon \boldsymbol{\l}_\varepsilon = - \int_\varepsilon  \sum_{b,\bk\in S^{b}(\varepsilon)}\boldsymbol{\l}_{\bk,b}
			\end{cases}  \nn
		\end{eqnarray}
		The solution is found simply by taking $\nabla_{\bf E}\alpha = 0$ and
		\begin{equation} \label{l_eps}
			\boldsymbol{\l}_\varepsilon =  - \frac{1}{\mathcal{N}(\varepsilon) }\sum_{b,\bk\in S^{b}(\varepsilon)}\boldsymbol{\l}_{\bk,b}.
		\end{equation}
		We conclude that the chemical potential does not change with the field while the charges get redistributed within each energy-shell {\it independently}. Such behavior was expected since the scattering is elastic and the entropy is additive on each energy shell. As a consequence also the energy density is left invariant by the field.

		\section{Symmetries} \label{App: Syms}

		\subsection{Role of space-group symmetries}\label{sec: SpaceSym}
		Space group symmetries of a system will manifest in transport if also the disorder has on average the same symmetry.
		
		A clean system with space group symmetries has the property $ H(R{\bf k}) = U^R\,H({\bf k})U^{R\dagger}$, where $R$ is the representation of the space group element acting on momenta while $U^R$ is the representation upon the orbital space. The symmetry implies that for all $b,k$ it exists a $b'$ such that $ P_{R\bk,b'} = U^R\, P_{\bk,b}\,U^{R\dagger}$. The disorder may have the same simmetry. There would be a "strong" symmetry if for each impurity $ W = U^R\,W U^{R\dagger}$ while only a "weak" one (or better, a statistical one) if, using trial matrices $X$ and $Y$,
		\begin{equation} \label{App:dis_sym}
		    X \circ D \circ Y= \left(U^R X U^{R\dagger} \right)\circ D \circ \left(U^R X U^{R\dagger} \right).
		\end{equation}  Luckily this difference is irrelevant here since all transport quantities depend on $W$ through $D$.
		
		Concerning disorder-independent band properties, energies stay constant over momenta along the same group-orbits; velocities obey $\bv_{R{\bf k},b'} = \bar R\bv_{{\bf k},b}$, with $\bar R$ the representation on real space of the group; the norms simply verify $v_{{R\bf k},b'} = v_{{\bf k},b}$. 
		
		Transport properties require that disorder be symmetric. Assuming it be the case, it is not difficult to show that the scattering times obey $\G_{R\bk,b'} = \G_{\bk,b}$. Instead, the matrix-valued vector ${\bf F}$ determining the vertex corrections must satisfy a compelling equality
		\begin{equation} \label{app:Fsym}
		{\bf F}_\varepsilon = U^R\, \left(\bar R\,{\bf F}_\varepsilon\right)\,U^{R\dagger}
		\end{equation}
		for all elements of the group. For this reason, it is likely that vertex corrections vanish for systems with a high degree of symmetry. As an example, in FeSe mirror symmetries with $U^R=\mathbb{1}$ selectively make vertex corrections vanishing for some pockets along specific directions increasing the anisotropic transport in the nematic phase\cite{Note1}.
		
		Similarly to $D$, the $K$ tensor has the property
		\begin{equation} \label{App:K_sym}
		    X \circ K \circ Y= \left(U^R X U^{R\dagger} \right)\circ K \circ \left(U^R Y U^{R\dagger} \right).
		\end{equation} 
		As a side remark, this invariance implies that the $\circ$-eigenvalues of both $K$ and $D$ are degenerate with multiplicity given by the dimensions of the irreducible representations of the group. 
		
		Finally one has the compelling constraint for the average renormalized velocities $\boldsymbol{\l}_\varepsilon = \bar R \boldsymbol{\l}_\varepsilon$ (see Eq. \eqref{l_eps}). Combining all constraints given by the symmetry, one finally find for the renormalized velocities ${\bf w}_{R\bk,b'} = \bar R\,{\bf w}_{\bk,b}$ while for dc conductivity tensor $\sigma = \bar R \sigma \bar R^T$.
		
		The IIE and the GUE are special among the other ones. Indeed, for all group symmetries $D^{(IIE)}$, being the $\circ$-identity, is always symmetric while for the GUE the independence of rates from momenta and the vanishing vertex corrections (see App. \ref{App: TK2}) lead always to symmetric transport quantities.

		\subsection{Time reversal symmetric models} \label{TRS}
		For time reversal invariant systems there exist an antiunitary operator such that $\mathcal T\,H(\bk)\,\mathcal T^{-1} =H(-\bk)$. One can express $\mathcal{T}=\hat U^{\mathcal T}\,\hat K$ with $\hat K$ the complex conjugation operator and $U^{\mathcal T}$ a unitary. Systems with broken time-reversal symmetry are described by version of the Boltzmann equation that include magnetic driving terms. We may understand our model, Eq. \eqref{boltz_eq}, as describing only those time-reversal symmetry broken systems where those driving terms lead to negligible effects.
		
		For what concerns band properties of time-reversal symmetric models, we have that for all $b,k$ it exists a $b'$ such that  $\xi_{\bk,b} = \xi_{-\bk,b'} ,\;\bv_{\bk,b} = - \bv_{-\bk,b'},  P_{\bk,b}  = \mathcal T P_{-\bk,b'}\mathcal T^{-1} \equiv U^{\mathcal T}P^*_{-\bk,b'}U^{{\mathcal T} \dagger}$. 
		In analogy with space-group symmetries (see Sec. \eqref{sec: SpaceSym}), if the disorder is time-reversal symmetric on average then the transport properties have a symmetry. Let's assume the disorder is symmetric i.e. for trial $X$ and $Y$  matrices, 
		\begin{equation} \label{Trev_invD} 
			X \circ D \circ Y= \left[\left(\mathcal T X \mathcal T^{-1} \right)\circ D \circ \left(\mathcal T Y \mathcal T^{-1}\right)\right]^*,
		\end{equation}
		where the last step follows by the definition of $D$. Observing that the product in Eq. \ref{Trev_invD} is positive definite when $X$ and $Y$ are projectors, it is easy to prove that $\Gamma_{\bk,b} = \Gamma_{-\bk,b'}$. Using the property $X \circ D \circ Y =Y\circ D^* \circ X$ one gets ${\bf F}_\varepsilon = - \mathcal T\,{\bf F}_\varepsilon\,\mathcal T^{-1} \equiv - U^{\mathcal T}\,{\bf F}^*_\varepsilon\,U^{{\mathcal T}\dagger}$. One can check that the invariance property in Eq. \eqref{Trev_invD} holds also for $K$. Quite surprisingly from these properties one finds $\boldsymbol{\l}_\varepsilon = 0$ for symmetric systems (see Eq. \eqref{l_eps}). It follows also ${\bf w}_{\bk,b} = - {\bf w}_{-\bk,b'}$ and a constraint for the integrand of $\sigma^{corr}_{ij}$ whose interpretation, however, is not physically transparent:
		$F^{i*}_{\varepsilon}   \circ D \circ \left(\mathbb{1} - K_{\varepsilon} \right)^{-1} \circ  F^j_{\varepsilon} =
		\left(U^{\mathcal T}\,F^{i}_{\varepsilon}\,U^{{\mathcal T}\dagger}\right) \circ  D \circ \left(\mathbb{1} - K_{\varepsilon} \right)^{-1} \circ \left(U^{\mathcal T}\,F^{j*}_{\varepsilon}\,U^{{\mathcal T}\dagger}\right)$.

		\section{Statistical ensembles for $D$} 
		
		Despite the lack of knowledge of the specific disorder configuration in a sample, one has often enough information about the character (structure, symmetry) of the disorder to opt for a statistical description using a specific matrix ensemble. In these paper we describe the ensembles related to time-reversal symmetry and orbital isotropy. Clearly other symmetries may be considered but may not be as universally relevant as these ones.

		\subsection{Wigner and IIE ensembles} \label{app: ensembles2}
		
		In the context of Random matrix Theory\cite{For10,Meh04} the Wigner ensembles consist of three ensembles GUE, GOE and GSE. They are vectorial spaces respectively of hermitian, symmetric real, symplectic self-dual (hermitian with quaternionic elements) matrices $W$ equipped with a Gaussian probabilistic measure. For our purpose we add a spatial label to the matrices $W_j$ and assume that for each impurity there is an independent distribution. Thus the measure of integration in Eq. \ref{Vprobs} is $
		{\mathcal D} W_c = C \prod_{m\geq m'}^{N_b} \mathrm{d} W_c^{mm'} \nn	$
		where $C$ is some normalization constant to ensure unit integral of the probability, $\mathrm{d} W_j^{mm'}$ stands for $\mathrm{d} W_j^{mm'} \,\mathrm{d} W_j^{*mm'}$ in the GUE case and $\prod_{q'=0,1}\mathrm{d} W_j^{(m0)(m'q')} \,\mathrm{d} W_j^{*(m0)(m'q')}$ in the GSE case with the second indices in the brackets labelling the quaternionic degree of freedom (cf. Eq. \ref{Tstats}). Notice that intra-orbital matrix elements are statistically greater than the inter-orbital one as the Gaussian in Eq. \ref{Vprobs2} can be written as $\exp\{-\left(\sum_m^{N_b} |W_{mm}|^2 + 2\sum_{m>m'}^{N_b} |  W^{mm'}|^2\right)/2\}$.
		
		These ensembles refer to a precise symmetry of each disorder configuration: spinless and time-reversal symmetric (GOE), spinfull and time-reversal symmetric (GSE)  or having no symmetries at all (GUE). This means $U^{\mathcal T}=\mathbb{1}$ in the GOE case and $U^{\mathcal T} = i\sigma_y$ acting on the spin sector in the GSE case (see App. \ref{TRS} for the definition of the time-reversal symmetry). 
		A generalization to ensembles with different $U^{\mathcal T}$ is trivial. Indeed in these cases there is always a unitary transformation that brings such systems to either a GOE-type time reversal or a GSE-type one\cite{For10}, and the invariance formula for $D$ can be easily written down using such unitary.
		
		We stress that at fixed disorder symmetry there is an infinite number of ensembles that are orbital-isotropic. Their distribution of eigenvectors of $W$ has to be uniformly distributed (Haar measure) while that of the eigenvalues must be of the form $\prod_{i>j}^{N_b} |E_i - E_j|^\beta\prod_{i}^{N_b}f(E_i)$ where the first term is the level repulsion function (from the Vandermonde determinant) whose strength is given by some positive number $\beta$ and $f(\cdot)$ is an arbitrary function fast-decaying at infinity. Notice that the GUE, the GOE and the GSE have $\beta=2,1,4$ respectively and $f(\cdot)$ is a simple Gaussian. Since in our treatment only products of the form $WW$ appear through $D$ (see for instance Eq. \ref{dis1imp}), only the second moment of the distribution is relevant, thus justifying the choice of the Wigner ensembles over more complex ones.

		By "Isotropic Intra-obital Ensemble" (IIE) we denote i.i.d. matrices of the form $W_c = v_c\mathbb{1}$ and a measure ${\mathcal D} V_c =  \mathrm{d} v_c$ for real numbers $v_c$. The IIE is the most direct generalization of the single band disorder. Actually considered the formulas we use, there is no even need to give the $v_c$s such a Gaussian probability and taking a fixed value $v_c = v_I$ would produce identical results (see for instance the disorder treatment in Ref. \cite{Bros16}). Also this ensemble is eigenvector-isotropic but in a trivial way and describes the disorder induced by electric-field fluctuations (as it equally couples to all orbitals without mixing them, in effect it couples to the total electronic charge). For this ensemble $\beta=0$. 
		
		\subsection{Derivation of Eq. \ref{Tstats}} \label{app: ensembles3}
		To obtain the elements of the tensor $D$, we simply had to evaluate Eq. \ref{Vprobs} with $\mathO = W^{mm'}_jW^{*nn'}_i$ at $i=j$. Notice that for the IIE and GUE the evaluation is very simple. We show here only the derivation of the GOE and GSE:
		
		GOE: In the case $m=m'$ then $n=n'=m$ must follow in order to have non-zero average. The gaussian average yields simply $\delta_{mm'}\delta_{nn'}\delta_{nm}$. In the case $m\neq m'$ then either $n=m$, and $n'=m'$ would follow, or $n=m'$, and $n'=m$ would follow since $W$ is symmetric. The first istance yields average $(1/2)\delta_{mn}\delta_{m'n'}(1-\delta_{mm'})$ where the last factor is to avoid double counting of diagonal case above. Similarly the second instance yields $(1/2)\delta_{mn'}\delta_{m'n}(1-\delta_{mm'})$. Summing all together we get the GOE expression in Eq. \ref{Tstats}.
		
		GSE: The diagonal element of $W$ consists in the quaternion $a_s\sigma_0$ with $a_s$ real $(s=1,\dots,N_b)$. These terms contribute to the average of $W^{(mq)(m'q')}W^{*(nr)(n'r')}$ as $ \delta_{mm'}\delta_{nn'}\delta_{nm}\delta_{qq'}\delta_{rr'}/2$. The off-diagonal term at position $ss'$ is a quaternion whose matrix representation is $((a,b),(-b^*,a^*))_{ss'}$. This quaternion has non trivial average when it is coupled to itself or to the quaternion at $s's$, which is its conjugate due to the hermiticity of $V$. In the first case the contribution will be $\delta_{mn}\delta_{m'n'}\delta_{qr}\delta_{q'r'}(1-\delta_{mm'})/2$, in the second case $(-1)^{q+q'}\delta_{mn'}\delta_{m'n}\delta_{q (1-r)}\delta_{q'(1-r')}(1-\delta_{mm'})/2$ where the first factor comes from the $(-b^*)$ element in the definition of a quaternion. Summing all together and observing that few terms cancel out we get the GSE expression in Eq. \ref{Tstats}.
		
		W.r.t. the $\circ$-product, $D_{GOE}>0$ while $D_{GSE}$ has indefinite signature.
		
		\subsection{Comparison between ensembles: the value of $\sigma_W$} \label{App:ensComparison}
		For each ensemble the variance $\sigma_W$ sets the average strength of the perturbation and does play the same role of $v_I$. 
		There is a problem when trying to compare different ensembles. Since the ensembles have a different number of degrees of freedom, it is somewhat arbitrary to claim that fixing equal values of $v_I$ and $\sigma_W$ for all ensembles is equivalent to set the disorders to {\it equal strength}. Such ambiguity is discussed here. Depending on the needed application, we suggest here two options to fix consistently the value of $\sigma_W$ for the different ensembles, at fixed equal value of $v_I$.
		
		1) The simplest choice is
		\be \label{App:choice1}
		\sigma_W=1 \quad \quad \quad \mathrm{all\;ensembles}
		\ee
		whereby all degrees of freedom of $W$, irrespectively of the ensemble, have the same variance modulo their multiplicity i.e. $\langle d^2\rangle = 1/m_d$, for a generic free variable $d$ appearing with multiplicity $m_d$ in $W$.
		Such choice may be driven also by the mathematics of the rates and vertex corrections formulas. Since $D$ appears always contracted according to the $\circ$-product, one may want to force the sum of the $\circ$-eigenvalues of $D$ to equal the same value, say $N_b$, for all ensembles - the specific value $N_b$ is suggested by Wigner's semicircular law, by which eigenvalues of large matrices of size $N$ scale like $\sqrt{N}$\cite{For10,Meh04}. One can verify that this choice of scaling indeed accommodates such requirement. Such choice may be preferable if one studies transitions between ensembles where an external field (e.g. a magnetic field) activates some disorder degrees of freedom without affecting the preexisting ones\cite{Bee97}.
		
		2) The option above creates an unbalance of the total strength of disorder between ensembles that have more non-vanishing elements in $W$ than others. Therefore, for some applications, one may prefer to give $v_I$ the meaning of average strength per degree of freedom of $W$, disregarding whether it is allowed or not by symmetry. Given that the number of degrees of freedom of the symmetry-unconstrained (hermitian matrix) $W$ is $N_b^2$, for each ensemble we want to find $\sigma_W$ such that $\sum_{mm'} \langle|W^{mm'}|^2\rangle_W/N_b^2 \equiv \sum_{mm'} D^{mm'}_{mm'}/N_b^2 = 1$. Using Eq. \eqref{Tstats} one finds
		\begin{eqnarray} \label{App:choice2}
			\sigma_W = 	&=&\begin{cases}
				N_b & \quad\quad\quad\mathrm{IIE}\\
				1 & \quad\quad\;\mathrm{GUE} \\
				\sqrt{2N_b/(N_b+1)} & \quad\quad\;\mathrm{GOE} \\
				\sqrt{2N_b/(N_b-1)} & \quad\quad\;\,\mathrm{GSE} 
			\end{cases}
		\end{eqnarray}
		where $N_b$ includes the spin degree of freedom in the GSE case (so $N_b^{GSE}\geq 2$). 
		Notice how $\sigma_W$ diverges with increasing $N_b$ in the IIE case. This divergence is there to compensate for the small number of non-vanishing elements in the disorder matrix (located only along the diagonal).
		
		\section{Properties of operators and tensors}
		\subsection{Structure of $Q\t$ and relation to the tensors $K_\varepsilon$} \label{App: TK1}
		Because of the delta function in energy in $Q\t$ (see Eq. \ref{final_multi_Q}), the kernel may be seen as a block diagonal matrix coupling only states with same energy. We prove here that, at each $\varepsilon$, the non-vanishing eigenvalues at each $\varepsilon$-block are the same as $K_\varepsilon$. Thus such blocks are separable operators\cite{Kan97}. Moreover, we show explicitely the one-to-one correspondence between the eigenfunctions of $Q\t$ and the eigenvectors of $K_\varepsilon$. 
		
		Let $M_\varepsilon$ be a right eigenmatrix of $K_\varepsilon$ i.e. $K_\varepsilon \circ M_\varepsilon = \lambda_\varepsilon M_\varepsilon$ with $\lambda_\varepsilon$ a scalar. We define $a^\varepsilon_{\bk,b} = \;P^*_{\bk,b} \circ D \circ M_{\varepsilon}\,\delta(\varepsilon_{\bk,b} - \varepsilon)$. One has (we use here a compact notation for conciseness): 
		\begin{eqnarray}
			Q\t \; a^\varepsilon &=& P^*\circ D \circ P  \tau\; a^\varepsilon \nn \\
			&=& P^*\circ D \circ K_\varepsilon \circ M_\varepsilon = \lambda_\varepsilon P^*\circ D \circ M_\varepsilon \nn \\
			&=&\lambda_\varepsilon \; a^\varepsilon 
		\end{eqnarray} 
		Thus $a^\varepsilon$ is an eigenfunction of $Q\t$.
		Conversely, let $a_{\bk,b}$ be an eigenfunction of $Q\t$. Since $Q\t$ couples only momenta and bands with the same energies, $a_{\bk,b}$ has support only on an energy shell $\varepsilon$. Then we define $M_\varepsilon = \sum_{b,\bk\in S^b(\varepsilon)} (\frac{P\,\t\,a}{|\bv|} )_{\bk,b}$. When $M_\varepsilon\neq 0$, one has (we neglect here summations and labels),  
		\begin{eqnarray}
			K_\varepsilon\, M_\varepsilon&=& \frac{P \t P \circ D}{|\bv|} \circ P^*  \tau\; a^\varepsilon \nn \\
			&=& \frac{P \t}{|\bv|} Q\t \;a^\varepsilon = \lambda_\varepsilon \,\frac{P \t}{|\bv|} \;a^\varepsilon\nn \\
			&=&\lambda_\varepsilon \; M_\varepsilon. 
		\end{eqnarray}
		Thus, the correspondence of the eigenpairs of $Q\t$ and $K$ is proved. The separability is a consequence, but is also evident from Eq. \ref{final_multi_Q}. After selecting a specific energy shell the delta function is superfluous and the term $P^*_{\bk,b} \circ D \circ P_{\bk',b'}$ is a finite summation over ($N_b^2$) products of two functions ($P$) respectively of the left and right kernel indexes (momenta and band number), which is the definition of separability of an operator. 
		
		\subsection{Vanishing vertex corrections with GUE disorder}\label{App: TK2}
		With the GUE disorder rates depend only on energy, $\G^{b\,(GUE)}_{\bk} = \G^{(single)}$. This happens because $D$ has only one non-vanishing eigenvalue when eigen-decomposed w.r.t. the $\circ$ product and its associated eigen-matrix is the identity operator (remember that $\Tr P_{\bk,b} = 1$).
		
		Let's consider again the tensor $K_\varepsilon$, one has $(K_\varepsilon)^{mm'}_{nn'} = \kappa \;\sum_{b,\bk\in S^b(\varepsilon)} \left(P^{*mn} \,\frac{\t}{|\bv|}\right)_{\bk,b} \,\delta_{m'n'}$. The tensor structure carries over to the inverse $(1-K_\varepsilon)^{-1\,mm'}_{\quad nn'}  = N_{mn} \,\delta_{m'n'}$ for some $N$ matrix, leading to a vertex corrections proportional to the quantity (cf. Eq. \eqref{multi_w})
		\begin{equation} \label{vanishing}
			\sum_{b,\bk\in S^b(\varepsilon)} ({\bf v}/|{\bf v}|)_{\bk}^b = 0.
		\end{equation}
		To see why this term vanishes, we can replace the l.h.s with $\sum_{i} \int_{S_i} d \vec S$, where $i$ labels the Fermi surfaces and $\vec S$ is the normalized normal vector of the surface. Multiplying the quantity by an arbitrary constant vector $\vec a$, applying the divergence theorem to the integral and $\nabla \cdot \vec a = 0$ we obtain $\int_{S_i} d \vec S=0$ for each surface.
		
		\subsection{$\sigma^{corr}>0$ if $D$ is $\circ\,$-positive } \label{App: TK3}
		
		We have mentioned that $D$ is always positive while grouping together the indexes $(mm')$ and $(nn')$. When this happens, in a non-trivial way, also with the grouping $(mn)$ and $(m'n')$ i.e. the $\circ$ product, then the vertex corrections are stricly positive. In that case we can find $X$ such that $D =X^\dagger \circ X$ and redefine $P = X \circ P$ and $\tilde K = X \circ K \circ X^{-1}$. The tensor $\tilde K$ is manifestly positive w.r.t. the $\circ$ product, indeed if we see it as a matrix ${\tilde K}_\varepsilon^{(mn)(m'n')} = \kappa \;\sum_{b,\bk\in S^b(\varepsilon)} \left( {\tilde P}^{mn}\,\frac{\t}{|\bv|} \,\tilde P^{*m'n'}\right)_{\bk,b}$, being a sum of positive $1$-rank matrices. Defining $\tilde {\bf F} = X \circ {\bf F}$, the vertex correction contribution to the dc conductivity reads as $\sigma^{corr} = \kappa \; \tilde {\bf F}^*_{\varepsilon} \circ \left(\mathbb{1} - \tilde K_{\varepsilon} \right)^{-1} \circ  \tilde {\bf F}_{\varepsilon}$ which is clearly positive at all energies since $(1-\tilde K)^{-1}$ is a positive matrix.

		\section{Proofs of the claims in Sec. \ref{sec:dis_eng}} \label{App:dis_eng_proofs}

		We show here the proofs of Sec. \ref{sec:dis_eng}. Consider Eq. \eqref{gamma_def}. To prove claim "i)", we must find $D$ such that $ D \circ  M_\varepsilon \propto \mathbb{1}$, with $M_\varepsilon = \sum_{b',\bk'\in S(\varepsilon^{b'}_\bk)} \left(P/|\bv|\right)_{\bk',b'}$, so that the rates will be constant (or equivalently isotropic) on the shell $\varepsilon$. Suppose we have only one impurity type then using Eq. \eqref{dis1imp} and the definition of $\circ$-product we have $ D \circ  M_\varepsilon = W M_\varepsilon W^\dagger$. Since these matrices are Hermitian, we can choose $W$ to be diagonalized by the same unitary $O$ that diagonalizes $M_\varepsilon$ obtaining $W M_\varepsilon W^\dagger = O_\varepsilon m_\varepsilon|w|^2 O_\varepsilon\dagger$ with $m_\varepsilon,w$ the diagonal eigenvalue matrices. Fixing $|w|^2 = m^{-1}_\varepsilon$ ($M_\varepsilon$ is a positive matrix), we obtain the desired property. Notice that for such a system $\boldsymbol{\l}_{\varepsilon}=0$. However, despite the constant rates, vanishing vertex corrections are not a necessary consequence (as it is instead for systems with GUE disorder).
		
		To prove claim "ii)", we observe that we can write $\Gamma_{\bk,b} = \sum_i\sum_j m^j_\varepsilon p_i|e_{\bk,b} \cdot W_i \cdot O^j_\varepsilon|^2$ where the index $i$ runs over the different impurities types of the system appearing with probability $p_i$, $O^j_\varepsilon$ is the $j$-th eigenvector of $M_\varepsilon$ and $e_{\bk,b}$ an electronic eigenstate. 
		Clearly if $W_i \perp e_{\bk,b}$ that contribution to the sum will vanish. It is quite evident that, given the positivity of the sum in $p_i$, a less heterogeneous disorder is always more apt to make a rate to vanish. So we may restrict to the single-impurity case. If there is an index $\bar j$ for which $m_{\bar j}$ vanishes it is clear that a disorder $W \parallel O_{\bar j}$ will make {\it all} rates to vanish, however, such eventuality is unlikely to happen because it is equivalent to say that the eigenstates do not span the whole orbital space. So assuming all $m_j>0$, we can choose the impurity disorder $W$ to have the minimal possible rank, rank $1$, (rank $0$ would imply $W=0$ which trivialize the problem) and we can tune it to be orthogonal to at least $N_b-1$ electronic eigenvector at different ${\bk,b}$ points.

		A desired claim one may like to make about vertex corrections is that it exists always a tensor $D$ that makes them vanishing. As one may have expected such kind of statements are difficult to prove. We first notice that $F_\varepsilon^\alpha =0 (\alpha=x,y,z)$ does not hold in a generic non-symmetric system. Even more, there are systems where it is non-vanishing {\it independently} of the disorder. This can happen because nothing forbids that some non-diagonal element of the projectors (see definition of $\bf F$ Eq. \eqref{Fdef}), say $P_{ij}$, may have always the same sign of $v^\alpha$ varying $\bk$ and $b$. Then since all the other integrands in $\bf F$ (in particular the rates) are positive one cannot make the element $F^\alpha_{ij}$ to vanish by tuning $D$ in these systems. However since in the vertex correction term $\bf F$ appears in the term $D \circ \left(\mathbb{1} - K_{\varepsilon_\bk} \right)^{-1} \circ F^\alpha $ one can hope to tune $D$ and make the full term vanishing. The tensor $\left(\mathbb{1} - K_{\varepsilon_\bk} \right)^{-1}$ cannot have a zero $\circ$-eigenvalue by the properties of $K$ (see App. \ref{App: TK1}), then the term can vanish only if it exists a $D$ such that $F^i$ is an eigenmatrix of $D$ with vanishing $\circ$-eigenvalue. To determine this condition one has to solve a non-linear system in the elements of $D$ which is hard to analyze and may be solved only numerically. Thus the only conclusion we can safely draw is that for a generic system $D$ will not produce vanishing vertex corrections if it is non-singular.
		
	\end{appendix}
		
	\footnotetext[1]{M. Marciani, L. Benfatto, {\it in preparation}.}
	\footnotetext[2]{A short proof for the generic inelastic case of $Q\t\leq 1$ 
		can be found in Taylor P. L., Proc. R. Soc. Lond. A \textbf{275} pag. 200--208 (1963). Even though the author did not notice it, the same proof implies also $Q\t\geq -1$. The adaptation of this proof in 
		our multiband elastic case is trivial.}
	\footnotetext[3]{Since $\bf v$ is orthogonal to $\{w_L\}$ at each energy shell, one may verify that also ${\bf F}$ is orthogonal to the left null-eigenvector of $1-K$, thus making Eq. \eqref{multi_w} non divergent.}
	
	\bibliography{Literature.bib}
\end{document}

%% file: Laras_tex_library.tex
\newcommand{\be}{\begin{equation}}
	\newcommand{\ee}{\end{equation}}
\newcommand{\bea}{\begin{eqnarray}}
	\newcommand{\eea}{\end{eqnarray}}

\def\a{\alpha}

\def\e{\varepsilon}

\def\l{\lambda}

\def\t{\tau}

\def\s{\sigma}

\def\G{\Gamma}

\def\pd{\partial}

\def\bk{{\bf k}}

\def\bx{{\bf x}}

\def\bv{{\bf v}}

\def\nn{\nonumber}
\def\lb{\label}
\def\pref#1{(\ref{#1})}

\newcount\bozza \bozza=0
\ifnum\bozza=1
\newdimen\shift \shift=-2truecm
\def\lb#1{%
	{\label{#1}\rlap{\kern\shift{$\scriptstyle#1$}}}}
\else\def\lb#1{\label{#1}} \fi

%% file: main.bbl
\begin{thebibliography}{49}%
\makeatletter
\providecommand \@ifxundefined [1]{%
 \@ifx{#1\undefined}
}%
\providecommand \@ifnum [1]{%
 \ifnum #1\expandafter \@firstoftwo
 \else \expandafter \@secondoftwo
 \fi
}%
\providecommand \@ifx [1]{%
 \ifx #1\expandafter \@firstoftwo
 \else \expandafter \@secondoftwo
 \fi
}%
\providecommand \natexlab [1]{#1}%
\providecommand \enquote  [1]{``#1''}%
\providecommand \bibnamefont  [1]{#1}%
\providecommand \bibfnamefont [1]{#1}%
\providecommand \citenamefont [1]{#1}%
\providecommand \href@noop [0]{\@secondoftwo}%
\providecommand \href [0]{\begingroup \@sanitize@url \@href}%
\providecommand \@href[1]{\@@startlink{#1}\@@href}%
\providecommand \@@href[1]{\endgroup#1\@@endlink}%
\providecommand \@sanitize@url [0]{\catcode `\\12\catcode `\$12\catcode
  `\&12\catcode `\#12\catcode `\^12\catcode `\_12\catcode `\%12\relax}%
\providecommand \@@startlink[1]{}%
\providecommand \@@endlink[0]{}%
\providecommand \url  [0]{\begingroup\@sanitize@url \@url }%
\providecommand \@url [1]{\endgroup\@href {#1}{\urlprefix }}%
\providecommand \urlprefix  [0]{URL }%
\providecommand \Eprint [0]{\href }%
\providecommand \doibase [0]{http://dx.doi.org/}%
\providecommand \selectlanguage [0]{\@gobble}%
\providecommand \bibinfo  [0]{\@secondoftwo}%
\providecommand \bibfield  [0]{\@secondoftwo}%
\providecommand \translation [1]{[#1]}%
\providecommand \BibitemOpen [0]{}%
\providecommand \bibitemStop [0]{}%
\providecommand \bibitemNoStop [0]{.\EOS\space}%
\providecommand \EOS [0]{\spacefactor3000\relax}%
\providecommand \BibitemShut  [1]{\csname bibitem#1\endcsname}%
\let\auto@bib@innerbib\@empty
\bibitem [{\citenamefont {Das~Sarma}\ \emph {et~al.}(2011)\citenamefont
  {Das~Sarma}, \citenamefont {Adam}, \citenamefont {Hwang},\ and\ \citenamefont
  {Rossi}}]{Das2011}%
  \BibitemOpen
  \bibfield  {author} {\bibinfo {author} {\bibfnamefont {S.}~\bibnamefont
  {Das~Sarma}}, \bibinfo {author} {\bibfnamefont {S.}~\bibnamefont {Adam}},
  \bibinfo {author} {\bibfnamefont {E.~H.}\ \bibnamefont {Hwang}}, \ and\
  \bibinfo {author} {\bibfnamefont {E.}~\bibnamefont {Rossi}},\ }\href@noop {}
  {\bibfield  {journal} {\bibinfo  {journal} {Rev. Mod. Phys.}\ }\textbf
  {\bibinfo {volume} {83}},\ \bibinfo {pages} {407} (\bibinfo {year}
  {2011})}\BibitemShut {NoStop}%
\bibitem [{\citenamefont {Gariglio}\ \emph {et~al.}(2016)\citenamefont
  {Gariglio}, \citenamefont {Gabay},\ and\ \citenamefont {Triscone}}]{Gar16}%
  \BibitemOpen
  \bibfield  {author} {\bibinfo {author} {\bibfnamefont {S.}~\bibnamefont
  {Gariglio}}, \bibinfo {author} {\bibfnamefont {M.}~\bibnamefont {Gabay}}, \
  and\ \bibinfo {author} {\bibfnamefont {J.-M.}\ \bibnamefont {Triscone}},\
  }\href@noop {} {\bibfield  {journal} {\bibinfo  {journal} {APL Materials}\
  }\textbf {\bibinfo {volume} {4}},\ \bibinfo {pages} {060701} (\bibinfo {year}
  {2016})}\BibitemShut {NoStop}%
\bibitem [{\citenamefont {Wang}\ \emph {et~al.}(2017)\citenamefont {Wang},
  \citenamefont {Lin}, \citenamefont {Wang}, \citenamefont {Yu},\ and\
  \citenamefont {Liao}}]{Wang2017}%
  \BibitemOpen
  \bibfield  {author} {\bibinfo {author} {\bibfnamefont {S.}~\bibnamefont
  {Wang}}, \bibinfo {author} {\bibfnamefont {B.-C.}\ \bibnamefont {Lin}},
  \bibinfo {author} {\bibfnamefont {A.-Q.}\ \bibnamefont {Wang}}, \bibinfo
  {author} {\bibfnamefont {D.-P.}\ \bibnamefont {Yu}}, \ and\ \bibinfo {author}
  {\bibfnamefont {Z.-M.}\ \bibnamefont {Liao}},\ }\href@noop {} {\bibfield
  {journal} {\bibinfo  {journal} {Advances in Physics: X}\ }\textbf {\bibinfo
  {volume} {2}},\ \bibinfo {pages} {518} (\bibinfo {year} {2017})}\BibitemShut
  {NoStop}%
\bibitem [{\citenamefont {Manzeli}\ \emph {et~al.}(2017)\citenamefont
  {Manzeli}, \citenamefont {Ovchinnikov}, \citenamefont {Pasquier},
  \citenamefont {Yazyev},\ and\ \citenamefont {Kis}}]{Man17}%
  \BibitemOpen
  \bibfield  {author} {\bibinfo {author} {\bibfnamefont {S.}~\bibnamefont
  {Manzeli}}, \bibinfo {author} {\bibfnamefont {D.}~\bibnamefont
  {Ovchinnikov}}, \bibinfo {author} {\bibfnamefont {D.}~\bibnamefont
  {Pasquier}}, \bibinfo {author} {\bibfnamefont {O.~V.}\ \bibnamefont
  {Yazyev}}, \ and\ \bibinfo {author} {\bibfnamefont {A.}~\bibnamefont {Kis}},\
  }\href@noop {} {\bibfield  {journal} {\bibinfo  {journal} {Nature Reviews
  Materials}\ }\textbf {\bibinfo {volume} {2}},\ \bibinfo {pages} {17033}
  (\bibinfo {year} {2017})}\BibitemShut {NoStop}%
\bibitem [{\citenamefont {Fernandes}\ and\ \citenamefont
  {Chubukov}(2016)}]{Fer16}%
  \BibitemOpen
  \bibfield  {author} {\bibinfo {author} {\bibfnamefont {R.~M.}\ \bibnamefont
  {Fernandes}}\ and\ \bibinfo {author} {\bibfnamefont {A.~V.}\ \bibnamefont
  {Chubukov}},\ }\href@noop {} {\bibfield  {journal} {\bibinfo  {journal}
  {Reports on Progress in Physics}\ }\textbf {\bibinfo {volume} {80}},\
  \bibinfo {pages} {014503} (\bibinfo {year} {2016})}\BibitemShut {NoStop}%
\bibitem [{\citenamefont {Gorbar}\ \emph {et~al.}(2018)\citenamefont {Gorbar},
  \citenamefont {Miransky}, \citenamefont {Shovkovy},\ and\ \citenamefont
  {Sukhachov}}]{Gor2018}%
  \BibitemOpen
  \bibfield  {author} {\bibinfo {author} {\bibfnamefont {E.}~\bibnamefont
  {Gorbar}}, \bibinfo {author} {\bibfnamefont {V.}~\bibnamefont {Miransky}},
  \bibinfo {author} {\bibfnamefont {I.}~\bibnamefont {Shovkovy}}, \ and\
  \bibinfo {author} {\bibfnamefont {P.}~\bibnamefont {Sukhachov}},\ }\href@noop
  {} {\bibfield  {journal} {\bibinfo  {journal} {Low Temperature Physics}\
  }\textbf {\bibinfo {volume} {44}},\ \bibinfo {pages} {487} (\bibinfo {year}
  {2018})}\BibitemShut {NoStop}%
\bibitem [{\citenamefont {Sangwan}\ and\ \citenamefont
  {Hersam}(2018)}]{San2018}%
  \BibitemOpen
  \bibfield  {author} {\bibinfo {author} {\bibfnamefont {V.~K.}\ \bibnamefont
  {Sangwan}}\ and\ \bibinfo {author} {\bibfnamefont {M.~C.}\ \bibnamefont
  {Hersam}},\ }\href@noop {} {\bibfield  {journal} {\bibinfo  {journal} {Annual
  Review of Physical Chemistry}\ }\textbf {\bibinfo {volume} {69}},\ \bibinfo
  {pages} {299} (\bibinfo {year} {2018})}\BibitemShut {NoStop}%
\bibitem [{\citenamefont {Gilmutdinov}\ \emph {et~al.}(2021)\citenamefont
  {Gilmutdinov}, \citenamefont {Vignolles}, \citenamefont {Proust},
  \citenamefont {Mukhamedshin}, \citenamefont {Balicas},\ and\ \citenamefont
  {Alloul}}]{Gil21}%
  \BibitemOpen
  \bibfield  {author} {\bibinfo {author} {\bibfnamefont {R.}~\bibnamefont
  {Gilmutdinov}, \bibfnamefont {I.~F.and~Sch\:onemann}}, \bibinfo {author}
  {\bibfnamefont {D.}~\bibnamefont {Vignolles}}, \bibinfo {author}
  {\bibfnamefont {C.}~\bibnamefont {Proust}}, \bibinfo {author} {\bibfnamefont
  {I.~R.}\ \bibnamefont {Mukhamedshin}}, \bibinfo {author} {\bibfnamefont
  {L.}~\bibnamefont {Balicas}}, \ and\ \bibinfo {author} {\bibfnamefont
  {H.}~\bibnamefont {Alloul}},\ }\href@noop {} {\bibfield  {journal} {\bibinfo
  {journal} {arXiv:2101.05252}\ } (\bibinfo {year} {2021})}\BibitemShut
  {NoStop}%
\bibitem [{\citenamefont {Ziman}(1960)}]{Ziman60}%
  \BibitemOpen
  \bibfield  {author} {\bibinfo {author} {\bibfnamefont {J.~M.}\ \bibnamefont
  {Ziman}},\ }\href@noop {} {\emph {\bibinfo {title} {{Electrons and phonons:
  the theory of transport phenomena in solids}}}},\ International series of
  monographs on physics\ (\bibinfo  {publisher} {Clarendon Press},\ \bibinfo
  {address} {Oxford},\ \bibinfo {year} {1960})\BibitemShut {NoStop}%
\bibitem [{\citenamefont {Mahan}(2000)}]{Mah00}%
  \BibitemOpen
  \bibfield  {author} {\bibinfo {author} {\bibfnamefont {G.~D.}\ \bibnamefont
  {Mahan}},\ }\href@noop {} {\emph {\bibinfo {title} {Many Particle Physics,
  Third Edition}}}\ (\bibinfo  {publisher} {Plenum},\ \bibinfo {address} {New
  York},\ \bibinfo {year} {2000})\BibitemShut {NoStop}%
\bibitem [{\citenamefont {Kamenev}(2011)}]{Kam2011}%
  \BibitemOpen
  \bibfield  {author} {\bibinfo {author} {\bibfnamefont {A.}~\bibnamefont
  {Kamenev}},\ }\href {\doibase 10.1017/CBO9781139003667} {\emph {\bibinfo
  {title} {Field Theory of Non-Equilibrium Systems}}}\ (\bibinfo  {publisher}
  {Cambridge University Press},\ \bibinfo {year} {2011})\BibitemShut {NoStop}%
\bibitem [{\citenamefont {Efetov}(2012)}]{Efe97}%
  \BibitemOpen
  \bibfield  {author} {\bibinfo {author} {\bibfnamefont {K.}~\bibnamefont
  {Efetov}},\ }\href@noop {} {\emph {\bibinfo {title} {{Supersymmetry in
  disorder and chaos}}}}\ (\bibinfo  {publisher} {Cambridge Univ. Press},\
  \bibinfo {address} {Cambridge, UK},\ \bibinfo {year} {2012})\BibitemShut
  {NoStop}%
\bibitem [{\citenamefont {Culcer}\ \emph {et~al.}(2017)\citenamefont {Culcer},
  \citenamefont {Sekine},\ and\ \citenamefont {Macdonald}}]{Cul2017}%
  \BibitemOpen
  \bibfield  {author} {\bibinfo {author} {\bibfnamefont {D.}~\bibnamefont
  {Culcer}}, \bibinfo {author} {\bibfnamefont {A.}~\bibnamefont {Sekine}}, \
  and\ \bibinfo {author} {\bibfnamefont {A.~H.}\ \bibnamefont {Macdonald}},\
  }\href {\doibase 10.1103/PhysRevB.96.035106} {\bibfield  {journal} {\bibinfo
  {journal} {Phys. Rev. B}\ }\textbf {\bibinfo {volume} {96}},\ \bibinfo
  {pages} {035106} (\bibinfo {year} {2017})}\BibitemShut {NoStop}%
\bibitem [{\citenamefont {Sekine}\ \emph {et~al.}(2017)\citenamefont {Sekine},
  \citenamefont {Culcer},\ and\ \citenamefont {MacDonald}}]{Sek2017}%
  \BibitemOpen
  \bibfield  {author} {\bibinfo {author} {\bibfnamefont {A.}~\bibnamefont
  {Sekine}}, \bibinfo {author} {\bibfnamefont {D.}~\bibnamefont {Culcer}}, \
  and\ \bibinfo {author} {\bibfnamefont {A.~H.}\ \bibnamefont {MacDonald}},\
  }\href@noop {} {\bibfield  {journal} {\bibinfo  {journal} {Phys. Rev. B}\
  }\textbf {\bibinfo {volume} {96}},\ \bibinfo {pages} {235134} (\bibinfo
  {year} {2017})}\BibitemShut {NoStop}%
\bibitem [{\citenamefont {Xiao}\ \emph {et~al.}(2019)\citenamefont {Xiao},
  \citenamefont {Du},\ and\ \citenamefont {Niu}}]{Cong19}%
  \BibitemOpen
  \bibfield  {author} {\bibinfo {author} {\bibfnamefont {C.}~\bibnamefont
  {Xiao}}, \bibinfo {author} {\bibfnamefont {Z.~Z.}\ \bibnamefont {Du}}, \ and\
  \bibinfo {author} {\bibfnamefont {Q.}~\bibnamefont {Niu}},\ }\href {\doibase
  10.1103/PhysRevB.100.165422} {\bibfield  {journal} {\bibinfo  {journal}
  {Phys. Rev. B}\ }\textbf {\bibinfo {volume} {100}},\ \bibinfo {pages}
  {165422} (\bibinfo {year} {2019})}\BibitemShut {NoStop}%
\bibitem [{\citenamefont {Stedman}\ and\ \citenamefont {Woods}(2020)}]{Sted20}%
  \BibitemOpen
  \bibfield  {author} {\bibinfo {author} {\bibfnamefont {T.}~\bibnamefont
  {Stedman}}\ and\ \bibinfo {author} {\bibfnamefont {L.~M.}\ \bibnamefont
  {Woods}},\ }\href@noop {} {\bibfield  {journal} {\bibinfo  {journal}
  {Physical Review Research}\ }\textbf {\bibinfo {volume} {2}},\ \bibinfo
  {pages} {33086} (\bibinfo {year} {2020})}\BibitemShut {NoStop}%
\bibitem [{\citenamefont {Cepellotti}\ and\ \citenamefont
  {Kozinsky}(2021)}]{Cep21}%
  \BibitemOpen
  \bibfield  {author} {\bibinfo {author} {\bibfnamefont {A.}~\bibnamefont
  {Cepellotti}}\ and\ \bibinfo {author} {\bibfnamefont {B.}~\bibnamefont
  {Kozinsky}},\ }\href {\doibase https://doi.org/10.1016/j.mtphys.2021.100412}
  {\bibfield  {journal} {\bibinfo  {journal} {Materials Today Physics}\
  }\textbf {\bibinfo {volume} {19}},\ \bibinfo {pages} {100412} (\bibinfo
  {year} {2021})}\BibitemShut {NoStop}%
\bibitem [{\citenamefont {Faulkner}\ \emph {et~al.}(2010)\citenamefont
  {Faulkner}, \citenamefont {Iqbal}, \citenamefont {Liu}, \citenamefont
  {McGreevy},\ and\ \citenamefont {Vegh}}]{Fau10}%
  \BibitemOpen
  \bibfield  {author} {\bibinfo {author} {\bibfnamefont {T.}~\bibnamefont
  {Faulkner}}, \bibinfo {author} {\bibfnamefont {N.}~\bibnamefont {Iqbal}},
  \bibinfo {author} {\bibfnamefont {H.}~\bibnamefont {Liu}}, \bibinfo {author}
  {\bibfnamefont {J.}~\bibnamefont {McGreevy}}, \ and\ \bibinfo {author}
  {\bibfnamefont {D.}~\bibnamefont {Vegh}},\ }\href@noop {} {\bibfield
  {journal} {\bibinfo  {journal} {Science}\ }\textbf {\bibinfo {volume}
  {329}},\ \bibinfo {pages} {1043} (\bibinfo {year} {2010})}\BibitemShut
  {NoStop}%
\bibitem [{\citenamefont {Adams}\ and\ \citenamefont {Yaida}(2015)}]{Ada15}%
  \BibitemOpen
  \bibfield  {author} {\bibinfo {author} {\bibfnamefont {A.}~\bibnamefont
  {Adams}}\ and\ \bibinfo {author} {\bibfnamefont {S.}~\bibnamefont {Yaida}},\
  }\href@noop {} {\bibfield  {journal} {\bibinfo  {journal} {Phys. Rev. D}\
  }\textbf {\bibinfo {volume} {92}},\ \bibinfo {pages} {126008} (\bibinfo
  {year} {2015})}\BibitemShut {NoStop}%
\bibitem [{\citenamefont {Lucas}\ and\ \citenamefont {Sachdev}(2015)}]{Luc15}%
  \BibitemOpen
  \bibfield  {author} {\bibinfo {author} {\bibfnamefont {A.}~\bibnamefont
  {Lucas}}\ and\ \bibinfo {author} {\bibfnamefont {S.}~\bibnamefont
  {Sachdev}},\ }\href {\doibase
  https://doi.org/10.1016/j.nuclphysb.2015.01.017} {\bibfield  {journal}
  {\bibinfo  {journal} {Nuclear Physics B}\ }\textbf {\bibinfo {volume}
  {892}},\ \bibinfo {pages} {239} (\bibinfo {year} {2015})}\BibitemShut
  {NoStop}%
\bibitem [{\citenamefont {Schwab}\ and\ \citenamefont
  {Raimondi}(2002)}]{Sch02}%
  \BibitemOpen
  \bibfield  {author} {\bibinfo {author} {\bibfnamefont {P.}~\bibnamefont
  {Schwab}}\ and\ \bibinfo {author} {\bibfnamefont {R.}~\bibnamefont
  {Raimondi}},\ }\href@noop {} {\bibfield  {journal} {\bibinfo  {journal} {The
  European Physical Journal B - Condensed Matter and Complex Systems}\ }\textbf
  {\bibinfo {volume} {25}},\ \bibinfo {pages} {483} (\bibinfo {year}
  {2002})}\BibitemShut {NoStop}%
\bibitem [{\citenamefont {Brosco}\ \emph {et~al.}(2016)\citenamefont {Brosco},
  \citenamefont {Benfatto}, \citenamefont {Cappelluti},\ and\ \citenamefont
  {Grimaldi}}]{Bros16}%
  \BibitemOpen
  \bibfield  {author} {\bibinfo {author} {\bibfnamefont {V.}~\bibnamefont
  {Brosco}}, \bibinfo {author} {\bibfnamefont {L.}~\bibnamefont {Benfatto}},
  \bibinfo {author} {\bibfnamefont {E.}~\bibnamefont {Cappelluti}}, \ and\
  \bibinfo {author} {\bibfnamefont {C.}~\bibnamefont {Grimaldi}},\ }\href
  {\doibase 10.1103/PhysRevLett.116.166602} {\bibfield  {journal} {\bibinfo
  {journal} {Phys. Rev. Lett.}\ }\textbf {\bibinfo {volume} {116}},\ \bibinfo
  {pages} {166602} (\bibinfo {year} {2016})}\BibitemShut {NoStop}%
\bibitem [{\citenamefont {Allen}(1978)}]{All78}%
  \BibitemOpen
  \bibfield  {author} {\bibinfo {author} {\bibfnamefont {P.~B.}\ \bibnamefont
  {Allen}},\ }\href@noop {} {\bibfield  {journal} {\bibinfo  {journal} {Phys.
  Rev. B}\ }\textbf {\bibinfo {volume} {17}},\ \bibinfo {pages} {3725}
  (\bibinfo {year} {1978})}\BibitemShut {NoStop}%
\bibitem [{\citenamefont {Breitkreiz}\ \emph {et~al.}(2014)\citenamefont
  {Breitkreiz}, \citenamefont {Brydon},\ and\ \citenamefont {Timm}}]{Bre14}%
  \BibitemOpen
  \bibfield  {author} {\bibinfo {author} {\bibfnamefont {M.}~\bibnamefont
  {Breitkreiz}}, \bibinfo {author} {\bibfnamefont {P.~M.~R.}\ \bibnamefont
  {Brydon}}, \ and\ \bibinfo {author} {\bibfnamefont {C.}~\bibnamefont
  {Timm}},\ }\href@noop {} {\bibfield  {journal} {\bibinfo  {journal} {Phys.
  Rev. B}\ }\textbf {\bibinfo {volume} {89}},\ \bibinfo {pages} {245106}
  (\bibinfo {year} {2014})}\BibitemShut {NoStop}%
\bibitem [{\citenamefont {Bychkov}\ and\ \citenamefont
  {E.~I.~Rashba}(1984)}]{Byc84}%
  \BibitemOpen
  \bibfield  {author} {\bibinfo {author} {\bibfnamefont {Y.~A.}\ \bibnamefont
  {Bychkov}}\ and\ \bibinfo {author} {\bibfnamefont {P.}~\bibnamefont
  {E.~I.~Rashba}},\ }\href@noop {} {\bibfield  {journal} {\bibinfo  {journal}
  {Zh. Eksp. Teor. Fiz.}\ }\textbf {\bibinfo {volume} {39}} (\bibinfo {year}
  {1984 [JETP Lett. 39, 78 (1984)]})}\BibitemShut {NoStop}%
\bibitem [{\citenamefont {Gui}\ \emph {et~al.}(2004)\citenamefont {Gui},
  \citenamefont {Becker}, \citenamefont {Dai}, \citenamefont {Liu},
  \citenamefont {Qiu}, \citenamefont {Novik}, \citenamefont {Sch\"afer},
  \citenamefont {Shu}, \citenamefont {Chu}, \citenamefont {Buhmann},\ and\
  \citenamefont {Molenkamp}}]{Gui04}%
  \BibitemOpen
  \bibfield  {author} {\bibinfo {author} {\bibfnamefont {Y.~S.}\ \bibnamefont
  {Gui}}, \bibinfo {author} {\bibfnamefont {C.~R.}\ \bibnamefont {Becker}},
  \bibinfo {author} {\bibfnamefont {N.}~\bibnamefont {Dai}}, \bibinfo {author}
  {\bibfnamefont {J.}~\bibnamefont {Liu}}, \bibinfo {author} {\bibfnamefont
  {Z.~J.}\ \bibnamefont {Qiu}}, \bibinfo {author} {\bibfnamefont {E.~G.}\
  \bibnamefont {Novik}}, \bibinfo {author} {\bibfnamefont {M.}~\bibnamefont
  {Sch\"afer}}, \bibinfo {author} {\bibfnamefont {X.~Z.}\ \bibnamefont {Shu}},
  \bibinfo {author} {\bibfnamefont {J.~H.}\ \bibnamefont {Chu}}, \bibinfo
  {author} {\bibfnamefont {H.}~\bibnamefont {Buhmann}}, \ and\ \bibinfo
  {author} {\bibfnamefont {L.~W.}\ \bibnamefont {Molenkamp}},\ }\href@noop {}
  {\bibfield  {journal} {\bibinfo  {journal} {Phys. Rev. B}\ }\textbf {\bibinfo
  {volume} {70}},\ \bibinfo {pages} {115328} (\bibinfo {year}
  {2004})}\BibitemShut {NoStop}%
\bibitem [{\citenamefont {Eremeev}\ \emph {et~al.}(2012)\citenamefont
  {Eremeev}, \citenamefont {Nechaev}, \citenamefont {Koroteev}, \citenamefont
  {Echenique},\ and\ \citenamefont {Chulkov}}]{Ere12}%
  \BibitemOpen
  \bibfield  {author} {\bibinfo {author} {\bibfnamefont {S.~V.}\ \bibnamefont
  {Eremeev}}, \bibinfo {author} {\bibfnamefont {I.~A.}\ \bibnamefont
  {Nechaev}}, \bibinfo {author} {\bibfnamefont {Y.~M.}\ \bibnamefont
  {Koroteev}}, \bibinfo {author} {\bibfnamefont {P.~M.}\ \bibnamefont
  {Echenique}}, \ and\ \bibinfo {author} {\bibfnamefont {E.~V.}\ \bibnamefont
  {Chulkov}},\ }\href@noop {} {\bibfield  {journal} {\bibinfo  {journal} {Phys.
  Rev. Lett.}\ }\textbf {\bibinfo {volume} {108}},\ \bibinfo {pages} {246802}
  (\bibinfo {year} {2012})}\BibitemShut {NoStop}%
\bibitem [{\citenamefont {Bahramy}\ \emph {et~al.}(2012)\citenamefont
  {Bahramy}, \citenamefont {Yang}, \citenamefont {Arita},\ and\ \citenamefont
  {Nagaosa}}]{Bah12}%
  \BibitemOpen
  \bibfield  {author} {\bibinfo {author} {\bibfnamefont {M.~S.}\ \bibnamefont
  {Bahramy}}, \bibinfo {author} {\bibfnamefont {B.-J.}\ \bibnamefont {Yang}},
  \bibinfo {author} {\bibfnamefont {R.}~\bibnamefont {Arita}}, \ and\ \bibinfo
  {author} {\bibfnamefont {N.}~\bibnamefont {Nagaosa}},\ }\href@noop {}
  {\bibfield  {journal} {\bibinfo  {journal} {Nature Communications}\ }\textbf
  {\bibinfo {volume} {3}},\ \bibinfo {pages} {679} (\bibinfo {year}
  {2012})}\BibitemShut {NoStop}%
\bibitem [{\citenamefont {Ohtomo}\ and\ \citenamefont {Hwang}(2004)}]{Oht04}%
  \BibitemOpen
  \bibfield  {author} {\bibinfo {author} {\bibfnamefont {A.}~\bibnamefont
  {Ohtomo}}\ and\ \bibinfo {author} {\bibfnamefont {H.~Y.}\ \bibnamefont
  {Hwang}},\ }\href@noop {} {\bibfield  {journal} {\bibinfo  {journal}
  {Nature}\ }\textbf {\bibinfo {volume} {427}},\ \bibinfo {pages} {423}
  (\bibinfo {year} {2004})}\BibitemShut {NoStop}%
\bibitem [{\citenamefont {Caviglia}\ \emph {et~al.}(2008)\citenamefont
  {Caviglia}, \citenamefont {Gariglio}, \citenamefont {Reyren}, \citenamefont
  {Jaccard}, \citenamefont {Schneider}, \citenamefont {Gabay}, \citenamefont
  {Thiel}, \citenamefont {Hammerl}, \citenamefont {Mannhart},\ and\
  \citenamefont {Triscone}}]{Cav08}%
  \BibitemOpen
  \bibfield  {author} {\bibinfo {author} {\bibfnamefont {A.~D.}\ \bibnamefont
  {Caviglia}}, \bibinfo {author} {\bibfnamefont {S.}~\bibnamefont {Gariglio}},
  \bibinfo {author} {\bibfnamefont {N.}~\bibnamefont {Reyren}}, \bibinfo
  {author} {\bibfnamefont {D.}~\bibnamefont {Jaccard}}, \bibinfo {author}
  {\bibfnamefont {T.}~\bibnamefont {Schneider}}, \bibinfo {author}
  {\bibfnamefont {M.}~\bibnamefont {Gabay}}, \bibinfo {author} {\bibfnamefont
  {S.}~\bibnamefont {Thiel}}, \bibinfo {author} {\bibfnamefont
  {G.}~\bibnamefont {Hammerl}}, \bibinfo {author} {\bibfnamefont
  {J.}~\bibnamefont {Mannhart}}, \ and\ \bibinfo {author} {\bibfnamefont
  {J.-M.}\ \bibnamefont {Triscone}},\ }\href@noop {} {\bibfield  {journal}
  {\bibinfo  {journal} {Nature}\ }\textbf {\bibinfo {volume} {456}},\ \bibinfo
  {pages} {624} (\bibinfo {year} {2008})}\BibitemShut {NoStop}%
\bibitem [{\citenamefont {Kohn}\ and\ \citenamefont {Luttinger}(1957)}]{Koh57}%
  \BibitemOpen
  \bibfield  {author} {\bibinfo {author} {\bibfnamefont {W.}~\bibnamefont
  {Kohn}}\ and\ \bibinfo {author} {\bibfnamefont {J.~M.}\ \bibnamefont
  {Luttinger}},\ }\href@noop {} {\bibfield  {journal} {\bibinfo  {journal}
  {Phys. Rev.}\ }\textbf {\bibinfo {volume} {108}},\ \bibinfo {pages} {590}
  (\bibinfo {year} {1957})}\BibitemShut {NoStop}%
\bibitem [{\citenamefont {Luttinger}\ and\ \citenamefont {Kohn}(1958)}]{Lut58}%
  \BibitemOpen
  \bibfield  {author} {\bibinfo {author} {\bibfnamefont {J.~M.}\ \bibnamefont
  {Luttinger}}\ and\ \bibinfo {author} {\bibfnamefont {W.}~\bibnamefont
  {Kohn}},\ }\href@noop {} {\bibfield  {journal} {\bibinfo  {journal} {Phys.
  Rev.}\ }\textbf {\bibinfo {volume} {109}},\ \bibinfo {pages} {1892} (\bibinfo
  {year} {1958})}\BibitemShut {NoStop}%
\bibitem [{Note2()}]{Note2}%
  \BibitemOpen
  \bibinfo {note} {A short proof for the generic inelastic case of $Q\tau \leq
  1$ can be found in Taylor P. L., Proc. R. Soc. Lond. A \protect \textbf {275}
  pag. 200--208 (1963). Even though the author did not notice it, the same
  proof implies also $Q\tau \geq -1$. The adaptation of this proof in our
  multiband elastic case is trivial.}\BibitemShut {Stop}%
\bibitem [{\citenamefont {Forrester}(2010)}]{For10}%
  \BibitemOpen
  \bibfield  {author} {\bibinfo {author} {\bibfnamefont {P.~J.}\ \bibnamefont
  {Forrester}},\ }\href@noop {} {\emph {\bibinfo {title} {Log-Gases and Random
  Matrices (LMS-34)}}},\ London Mathematical Society Monographs\ (\bibinfo
  {publisher} {Princeton University Press},\ \bibinfo {year}
  {2010})\BibitemShut {NoStop}%
\bibitem [{\citenamefont {Schomerus}\ \emph {et~al.}(2015)\citenamefont
  {Schomerus}, \citenamefont {Marciani},\ and\ \citenamefont
  {Beenakker}}]{Sch15}%
  \BibitemOpen
  \bibfield  {author} {\bibinfo {author} {\bibfnamefont {H.}~\bibnamefont
  {Schomerus}}, \bibinfo {author} {\bibfnamefont {M.}~\bibnamefont {Marciani}},
  \ and\ \bibinfo {author} {\bibfnamefont {C.~W.~J.}\ \bibnamefont
  {Beenakker}},\ }\href@noop {} {\bibfield  {journal} {\bibinfo  {journal}
  {Phys. Rev. Lett.}\ }\textbf {\bibinfo {volume} {114}},\ \bibinfo {pages}
  {166803} (\bibinfo {year} {2015})}\BibitemShut {NoStop}%
\bibitem [{\citenamefont {Kane}\ and\ \citenamefont {Lubensky}(2014)}]{Kane14}%
  \BibitemOpen
  \bibfield  {author} {\bibinfo {author} {\bibfnamefont {C.~L.}\ \bibnamefont
  {Kane}}\ and\ \bibinfo {author} {\bibfnamefont {T.~C.}\ \bibnamefont
  {Lubensky}},\ }\href@noop {} {\bibfield  {journal} {\bibinfo  {journal}
  {Nature Physics}\ }\textbf {\bibinfo {volume} {10}},\ \bibinfo {pages} {39}
  (\bibinfo {year} {2014})}\BibitemShut {NoStop}%
\bibitem [{Note3()}]{Note3}%
  \BibitemOpen
  \bibinfo {note} {Since $\protect \bf v$ is orthogonal to $\{w_L\}$ at each
  energy shell, one may verify that also ${\protect \bf F}$ is orthogonal to
  the left null-eigenvector of $1-K$, thus making Eq. \protect \textup {\hbox
  {\mathsurround \z@ \protect \normalfont (\ignorespaces \ref {multi_w}\unskip
  \@@italiccorr )}} non divergent.}\BibitemShut {Stop}%
\bibitem [{\citenamefont {Han}\ and\ \citenamefont {Kim}(2018)}]{Han18}%
  \BibitemOpen
  \bibfield  {author} {\bibinfo {author} {\bibfnamefont {J.-H.}\ \bibnamefont
  {Han}}\ and\ \bibinfo {author} {\bibfnamefont {K.-S.}\ \bibnamefont {Kim}},\
  }\href {\doibase 10.1103/PhysRevB.97.214206} {\bibfield  {journal} {\bibinfo
  {journal} {Phys. Rev. B}\ }\textbf {\bibinfo {volume} {97}},\ \bibinfo
  {pages} {214206} (\bibinfo {year} {2018})}\BibitemShut {NoStop}%
\bibitem [{\citenamefont {Evers}\ and\ \citenamefont {Mirlin}(2008)}]{Mir08}%
  \BibitemOpen
  \bibfield  {author} {\bibinfo {author} {\bibfnamefont {F.}~\bibnamefont
  {Evers}}\ and\ \bibinfo {author} {\bibfnamefont {A.~D.}\ \bibnamefont
  {Mirlin}},\ }\href@noop {} {\bibfield  {journal} {\bibinfo  {journal} {Rev.
  Mod. Phys.}\ }\textbf {\bibinfo {volume} {80}},\ \bibinfo {pages} {1355}
  (\bibinfo {year} {2008})}\BibitemShut {NoStop}%
\bibitem [{\citenamefont {Nagaosa}\ \emph {et~al.}(2010)\citenamefont
  {Nagaosa}, \citenamefont {Sinova}, \citenamefont {Onoda}, \citenamefont
  {MacDonald},\ and\ \citenamefont {Ong}}]{Nag10}%
  \BibitemOpen
  \bibfield  {author} {\bibinfo {author} {\bibfnamefont {N.}~\bibnamefont
  {Nagaosa}}, \bibinfo {author} {\bibfnamefont {J.}~\bibnamefont {Sinova}},
  \bibinfo {author} {\bibfnamefont {S.}~\bibnamefont {Onoda}}, \bibinfo
  {author} {\bibfnamefont {A.~H.}\ \bibnamefont {MacDonald}}, \ and\ \bibinfo
  {author} {\bibfnamefont {N.~P.}\ \bibnamefont {Ong}},\ }\href@noop {}
  {\bibfield  {journal} {\bibinfo  {journal} {Reviews of modern physics}\
  }\textbf {\bibinfo {volume} {82}},\ \bibinfo {pages} {1539} (\bibinfo {year}
  {2010})}\BibitemShut {NoStop}%
\bibitem [{\citenamefont {Ashok}\ \emph {et~al.}(2014)\citenamefont {Ashok},
  \citenamefont {Chevallier}, \citenamefont {Johnson}, \citenamefont {Sopori},\
  and\ \citenamefont {Okushi}}]{Ash02}%
  \BibitemOpen
  \bibfield  {author} {\bibinfo {author} {\bibfnamefont {S.}~\bibnamefont
  {Ashok}}, \bibinfo {author} {\bibfnamefont {J.}~\bibnamefont {Chevallier}},
  \bibinfo {author} {\bibfnamefont {N.}~\bibnamefont {Johnson}}, \bibinfo
  {author} {\bibfnamefont {B.}~\bibnamefont {Sopori}}, \ and\ \bibinfo {author}
  {\bibfnamefont {H.}~\bibnamefont {Okushi}},\ }\href@noop {} {\emph {\bibinfo
  {title} {Defect- and Impurity-Engineered Semiconductors and Devices III}}}\
  (\bibinfo  {publisher} {Cambridge University Press},\ \bibinfo {address}
  {Cambridge, UK},\ \bibinfo {year} {2014})\BibitemShut {NoStop}%
\bibitem [{\citenamefont {Brandt}\ and\ \citenamefont {Ploog}(2006)}]{Bra06}%
  \BibitemOpen
  \bibfield  {author} {\bibinfo {author} {\bibfnamefont {O.}~\bibnamefont
  {Brandt}}\ and\ \bibinfo {author} {\bibfnamefont {K.~H.}\ \bibnamefont
  {Ploog}},\ }\href {\doibase 10.1038/nmat1728} {\bibfield  {journal} {\bibinfo
   {journal} {Nature Materials}\ }\textbf {\bibinfo {volume} {5}},\ \bibinfo
  {pages} {769} (\bibinfo {year} {2006})}\BibitemShut {NoStop}%
\bibitem [{\citenamefont {Raimondi}\ \emph {et~al.}(2012)\citenamefont
  {Raimondi}, \citenamefont {Schwab}, \citenamefont {Gorini},\ and\
  \citenamefont {Vignale}}]{Rai12}%
  \BibitemOpen
  \bibfield  {author} {\bibinfo {author} {\bibfnamefont {R.}~\bibnamefont
  {Raimondi}}, \bibinfo {author} {\bibfnamefont {P.}~\bibnamefont {Schwab}},
  \bibinfo {author} {\bibfnamefont {C.}~\bibnamefont {Gorini}}, \ and\ \bibinfo
  {author} {\bibfnamefont {G.}~\bibnamefont {Vignale}},\ }\href@noop {}
  {\bibfield  {journal} {\bibinfo  {journal} {Annalen der Physik}\ }\textbf
  {\bibinfo {volume} {524}} (\bibinfo {year} {2012})}\BibitemShut {NoStop}%
\bibitem [{\citenamefont {Xiao}\ \emph {et~al.}(2016)\citenamefont {Xiao},
  \citenamefont {Li},\ and\ \citenamefont {Ma}}]{Cong16}%
  \BibitemOpen
  \bibfield  {author} {\bibinfo {author} {\bibfnamefont {C.}~\bibnamefont
  {Xiao}}, \bibinfo {author} {\bibfnamefont {D.}~\bibnamefont {Li}}, \ and\
  \bibinfo {author} {\bibfnamefont {Z.}~\bibnamefont {Ma}},\ }\href {\doibase
  10.1103/PhysRevB.93.075150} {\bibfield  {journal} {\bibinfo  {journal} {Phys.
  Rev. B}\ }\textbf {\bibinfo {volume} {93}},\ \bibinfo {pages} {075150}
  (\bibinfo {year} {2016})}\BibitemShut {NoStop}%
\bibitem [{Note1()}]{Note1}%
  \BibitemOpen
  \bibinfo {note} {M. Marciani, L. Benfatto, {\protect \it in
  preparation}.}\BibitemShut {Stop}%
\bibitem [{\citenamefont {Mehta}(2004)}]{Meh04}%
  \BibitemOpen
  \bibfield  {author} {\bibinfo {author} {\bibfnamefont {M.~L.}\ \bibnamefont
  {Mehta}},\ }\href@noop {} {\emph {\bibinfo {title} {Random Matrices}}},\
  \bibinfo {edition} {3rd}\ ed.\ (\bibinfo  {publisher} {Academic Press},\
  \bibinfo {address} {New York},\ \bibinfo {year} {2004})\BibitemShut {NoStop}%
\bibitem [{\citenamefont {Nagaev}(2020)}]{Nag20}%
  \BibitemOpen
  \bibfield  {author} {\bibinfo {author} {\bibfnamefont {K.~E.}\ \bibnamefont
  {Nagaev}},\ }\href {\doibase https://doi.org/10.1002/pssr.201900536}
  {\bibfield  {journal} {\bibinfo  {journal} {physica status solidi (RRL) --
  Rapid Research Letters}\ }\textbf {\bibinfo {volume} {14}},\ \bibinfo {pages}
  {1900536} (\bibinfo {year} {2020})}\BibitemShut {NoStop}%
\bibitem [{\citenamefont {Beenakker}(1997)}]{Bee97}%
  \BibitemOpen
  \bibfield  {author} {\bibinfo {author} {\bibfnamefont {C.~W.~J.}\
  \bibnamefont {Beenakker}},\ }\href {\doibase 10.1103/RevModPhys.69.731}
  {\bibfield  {journal} {\bibinfo  {journal} {Rev. Mod. Phys.}\ }\textbf
  {\bibinfo {volume} {69}},\ \bibinfo {pages} {731} (\bibinfo {year}
  {1997})}\BibitemShut {NoStop}%
\bibitem [{\citenamefont {Kanwal}(1997)}]{Kan97}%
  \BibitemOpen
  \bibfield  {author} {\bibinfo {author} {\bibfnamefont {R.~P.}\ \bibnamefont
  {Kanwal}},\ }\href@noop {} {\emph {\bibinfo {title} {Integral Equations With
  Separable Kernels. In: Linear Integral Equations}}}\ (\bibinfo  {publisher}
  {Birkh{\"a}user},\ \bibinfo {address} {Boston{(MA)}},\ \bibinfo {year}
  {1997})\BibitemShut {NoStop}%
\end{thebibliography}%
